\documentclass[pra, aps, superscriptaddress, twocolumn, showpacs]{revtex4}
\usepackage{amsmath, amsfonts, amssymb, mathrsfs}
\usepackage{graphicx, subfigure, color}
\usepackage{bm}

\renewcommand{\d}{\partial}
\newcommand{\der}[2]{\frac{d#1}{d#2}}

\newcommand{\pder}[2]{\frac{\partial #1}{\partial #2}}

\newcommand{\lp}{\bigg(}
\newcommand{\rp}{\bigg)}
\newcommand{\lb}{\bigg[}
\newcommand{\rb}{\bigg]}

\renewcommand{\vec}[1]{\mathbf{#1}}
\newcommand{\vh}[1]{\hat{\vec{#1}}}

\newcommand{\vk}{\vec{k}}

\newcommand{\ket}[1]{|#1\rangle}
\newcommand{\bra}[1]{\langle#1|}

\newcommand{\mean}[1]{\langle#1\rangle}
\newcommand{\tr}{\operatorname{Tr}}

\newcommand{\be}{\begin{equation}}
\newcommand{\ee}{\end{equation}}

\begin{document}

\title{Scaling of entanglement entropy across Lifshitz transitions}

\author{Marlon~Rodney}
\affiliation{Department of Physics \& Astronomy, McMaster University, Hamilton, Ontario L8S 4M1, Canada}
\author{H.~Francis~Song}
\affiliation{Department of Physics, Yale University, New Haven, CT 06520}
\author{Sung-Sik~Lee}
\affiliation{Department of Physics \& Astronomy, McMaster University, Hamilton, Ontario L8S 4M1, Canada}
\affiliation{Perimeter Institute for Theoretical Physics, Waterloo, Ontario N2L 2Y5, Canada}
\author{Karyn~Le~Hur}
\affiliation{Centre de Physique Th\'{e}orique, Ecole Polytechnique, CNRS, 91128 Palaiseau Cedex, France}
\affiliation{Department of Physics, Yale University, New Haven, CT 06520}
\author{Erik S. S{\o}rensen}
\affiliation{Department of Physics \& Astronomy, McMaster University, Hamilton, Ontario L8S 4M1, Canada}

\begin{abstract}
We investigate the scaling of the bipartite entanglement entropy across Lifshitz quantum phase transitions, 
where the topology of the Fermi surface changes without any changes in symmetry. 
We present both numerical and analytical results which show that 
Lifshitz transitions are characterized by 
a well-defined set of critical exponents
for the entanglement entropy near the phase transition. 
In one dimension, we show that the entanglement entropy
exhibits a length scale that diverges as the system
approaches a Lifshitz critical point.
In two dimensions, 
the leading and sub-leading coefficients of the scaling of entanglement entropy 
show distinct power-law singularities at critical points.
The effect of weak interactions is investigated using the density matrix renormalization group algorithm.
\end{abstract}

\date{\today}
\maketitle

\section{Introduction} Quantum phase transitions are characterized by qualitative changes in the nature of the ground state as a parameter in the Hamiltonian is tuned across the transition, which mathematically manifests as non-analyticities in the ground-state energy and other local operators in the thermodynamic limit \cite{Sondhi97,Sachdev99}. 
The conventional framework for studying different phases of matter focuses on different symmetries and the identification of local order parameters that spontaneously break them.
However, the discovery of distinct phases of matter with the same symmetry
has recently introduced 
a more general notion of order associated
with the global topology of ground-state wave functions \cite{doi:10.1142/S0217979290000139}.
For example, 
different quantum Hall states \cite{Thouless82,Wen04} 
and topological insulators \cite{Kane05}
are characterized by topological invariants. 
Topologically ordered states  
have distinct patterns of long range entanglement
which can be probed using various measures of entanglement. 
In particular, one can consider a subsystem $\Omega$ of the full system with ground-state wave function $\ket{\Psi}$ and its associated reduced density matrix $\hat{\rho}_\Omega = \tr_{\overline{\Omega}}(\ket{\Psi}\bra{\Psi})$ obtained by tracing out the degrees of freedom in the remainder of the system, $\overline{\Omega}$. The bipartite entanglement entropy is defined as the von Neumann entropy of the reduced density matrix,
\be
  \mathcal{S}(\Omega) = -\tr(\hat{\rho}_\Omega \ln \hat{\rho}_\Omega),
\ee and is symmetric (for the pure state $\ket{\Psi}$) between $\Omega$ and $\overline{\Omega}$. 
Non-trivial topological order can be then detected from the universal component of the entanglement entropy \cite{Kitaev06,Levin06,Haque, Li,PhysRevB.84.075128,2012arXiv1205.4289J}.

As entanglement entropy can serve as a useful diagnostic for non-trivial gapped phases,
it can also be used to detect non-trivial topological structures in gapless phases.
Many interesting properties of the entanglement entropy, 
including striking universal scaling behavior in one-dimensional conformal field theories (CFTs), have been investigated over the years \cite{Holzhey94,Calabrese04}.
More recently, entanglement entropy has been computed for various gapless states in higher dimensions \cite{
PhysRevLett.96.181602,
PhysRevLett.97.050404,
Fradkin09,
PhysRevB.80.115122,
PhysRevLett.107.067202,
2011arXiv1112.1069S}. 
In this work, we use entanglement entropy to detect and characterize 
quantum phase transitions between two gapless phases with the same symmetry but with different topologies.
In particular, we study the scaling of the bipartite entanglement entropy across Lifshitz transitions, 
an example of a quantum phase transition in which there is no change in symmetry \cite{Lifshitz60,Blanter94,Yamaji06}. 
Instead, Lifshitz transitions are driven by changes in the topology of the Fermi surface, 
for example when two Fermi surfaces merge to form a single Fermi surface. 
The transition is signaled by van Hove singularities in the density of states and is due to the critical points where the Fermi velocity $d\varepsilon(\vk)/d\vk$ vanishes \emph{at} the Fermi surface. 
We give both analytical and numerical results for the scaling of entanglement entropy in some simple systems that possess Lifshitz transitions and 
show that the entanglement entropy exhibits critical behavior across such transitions. We also use the density matrix renormalization group (DMRG) algorithm to investigate the effect of weak interactions.

It is worth noting that, because of the equivalence 
between 
entanglement entropy and particle number fluctuations in free-fermion systems \cite{Klich09,Song11,Song12}, many of the conclusions regarding entanglement entropy in the systems studied below apply to particle number fluctuations. In addition, recent proposals for measuring the entanglement entropy in interacting systems \cite{Cardy11,Abanin12,Daley12,Nataf12} offer a promising way to observe the effects investigated in this work.

\section{Lifshitz transitions in 1D}

\begin{figure*}[t]
  \centering
  \subfigure[\ $t = 0.7$]{\includegraphics[width=5cm]{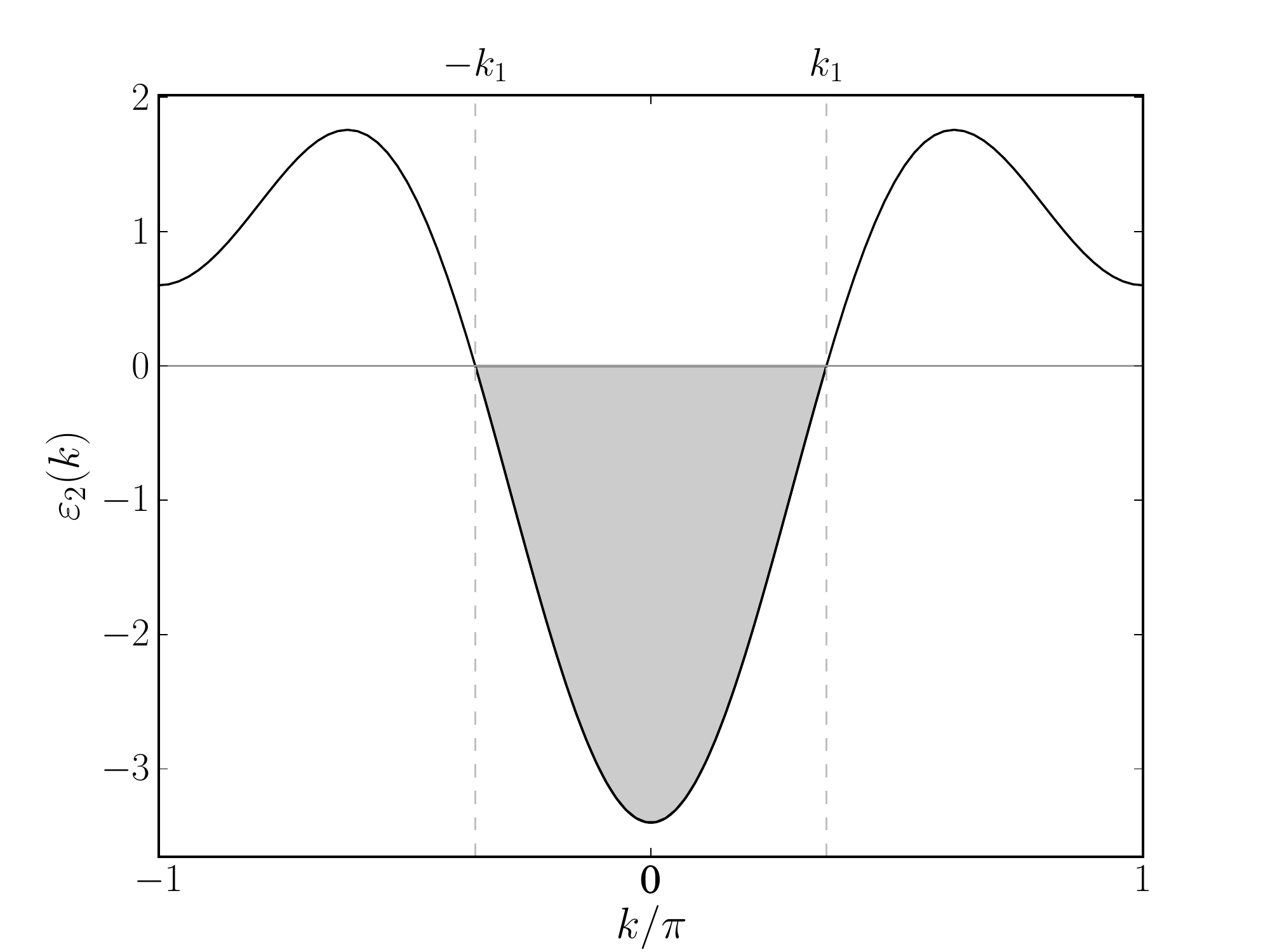}}
  \subfigure[\ $t = 1$]{\includegraphics[width=5cm]{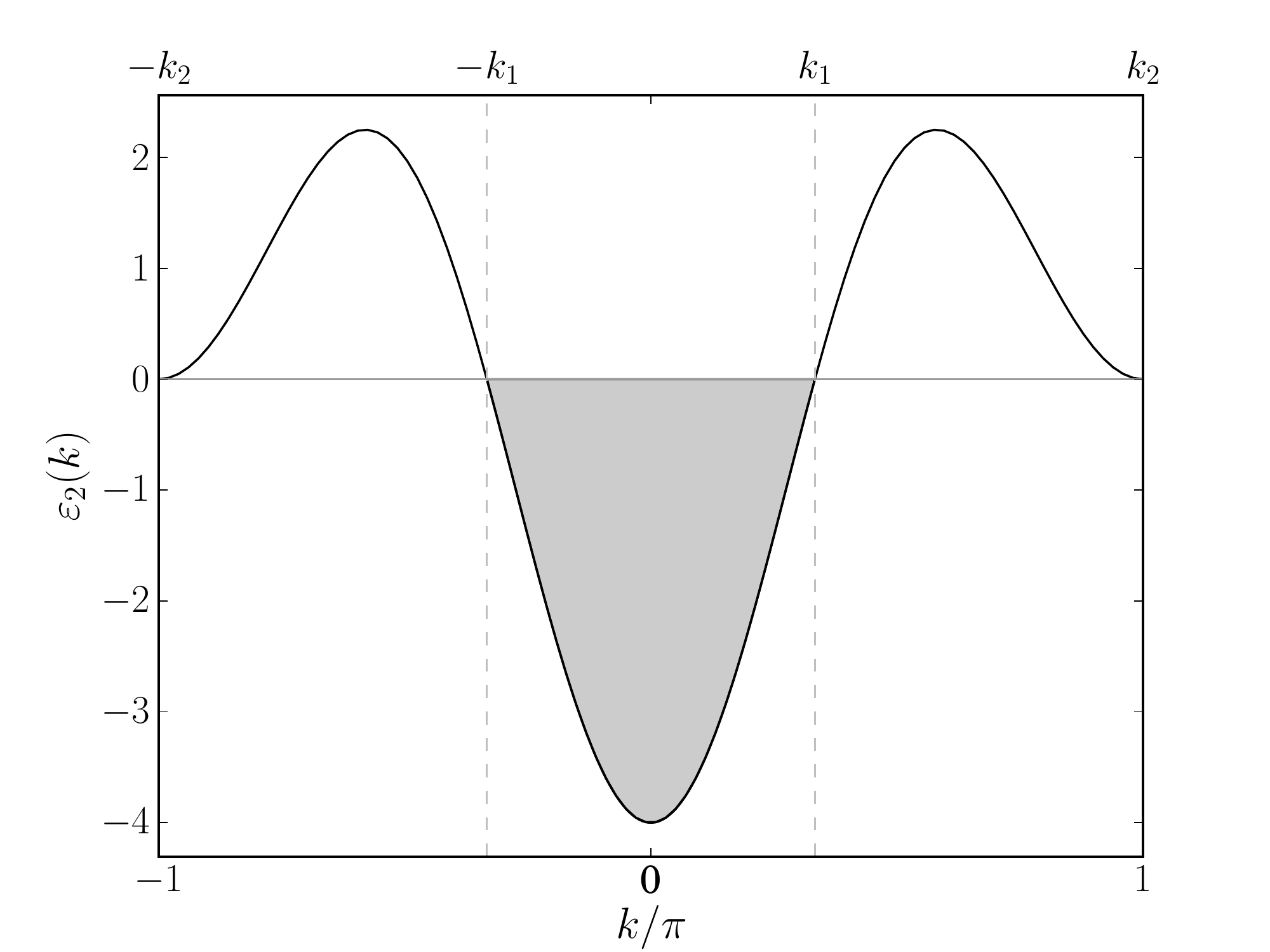}}
  \subfigure[\ $t = 1.3$]{\includegraphics[width=5cm]{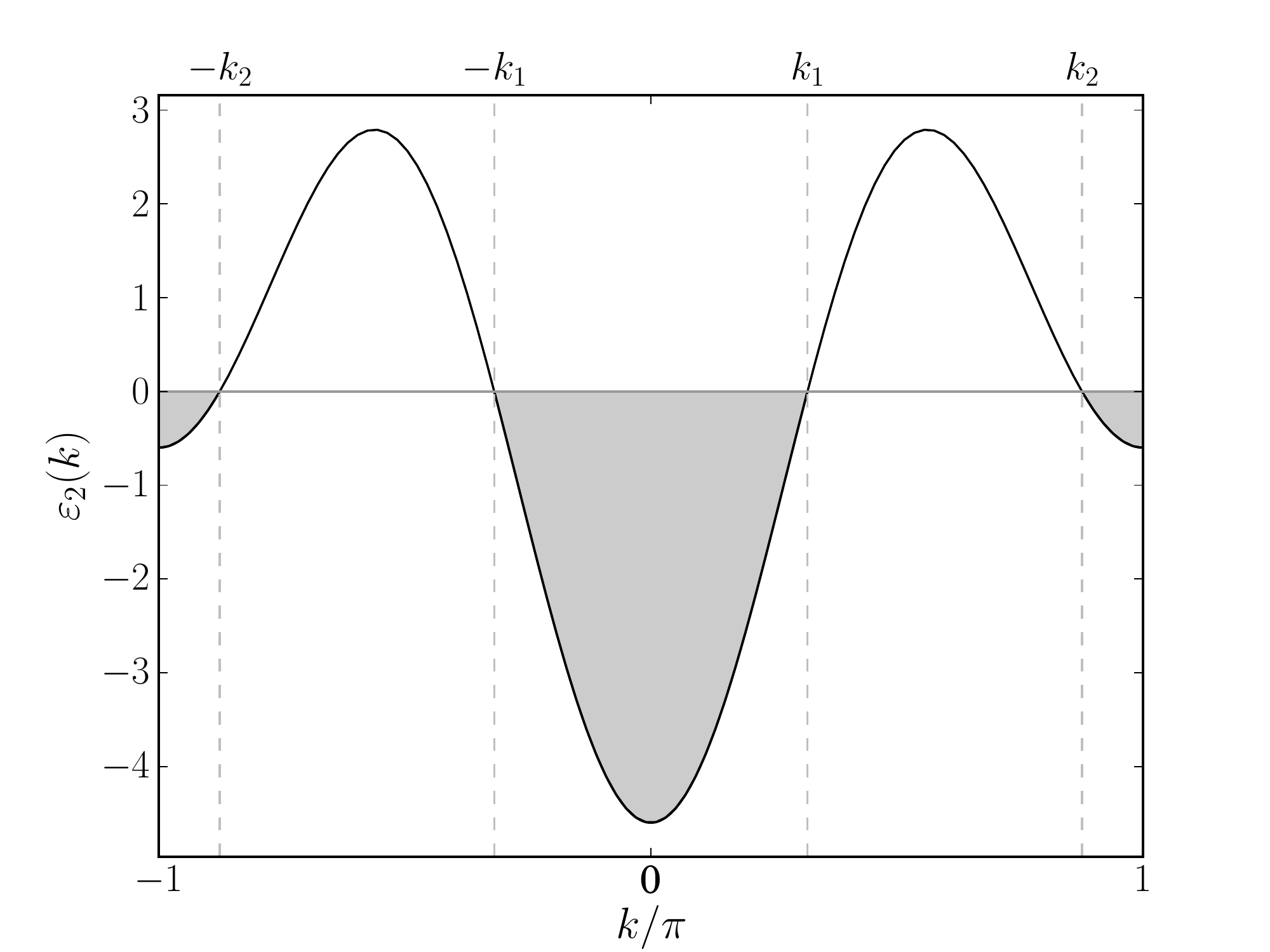}}
  \caption{(color online) Occupied modes (shaded) for the dispersion corresponding to Eq.~\eqref{eq:1d Hamiltonian} with $n=2$.}
  \label{fig:fs1d-n2}
\end{figure*}

\begin{figure*}[t]
  \centering
  \subfigure[\ $t = 0.1$]{\includegraphics[width=5cm]{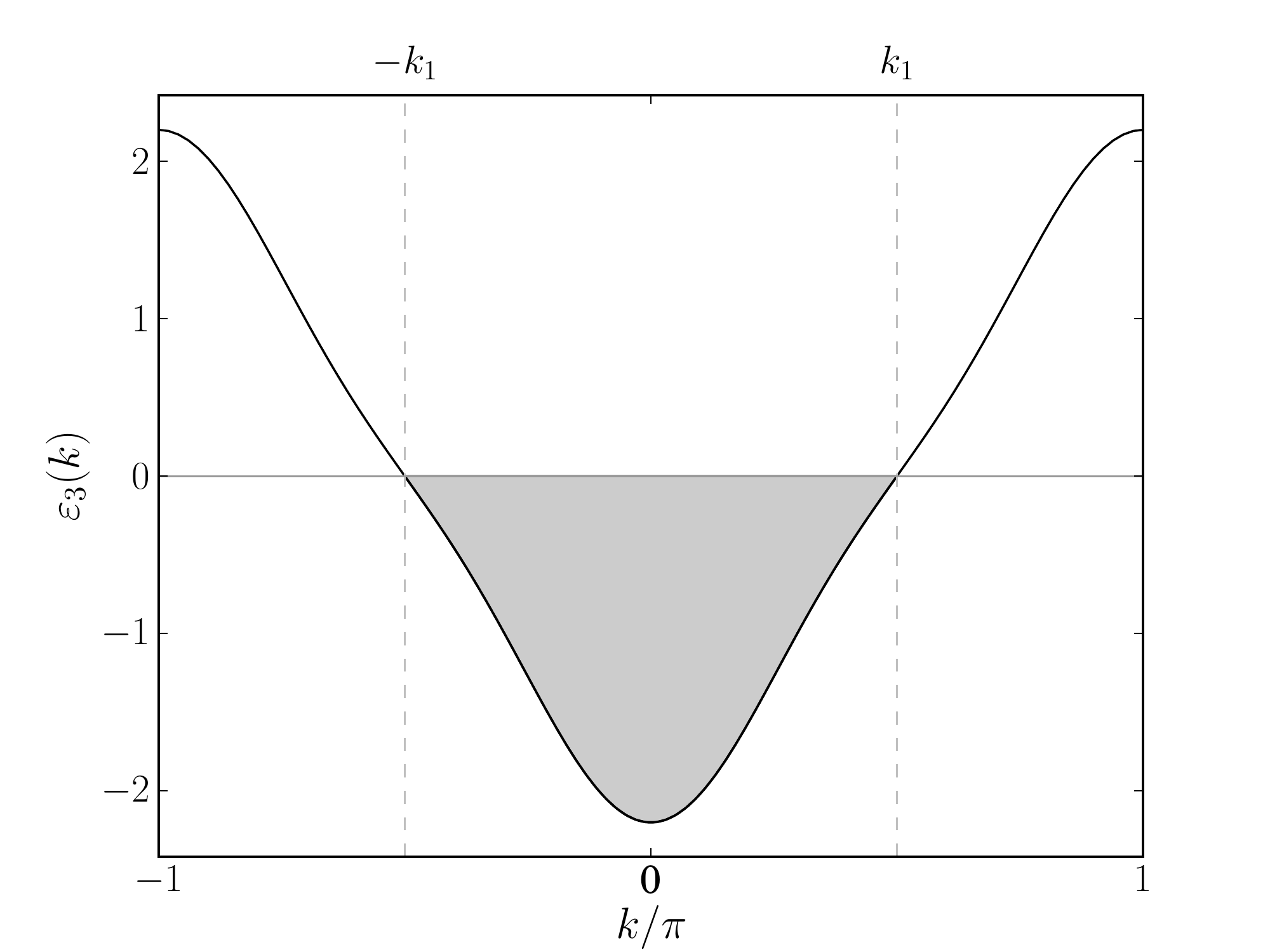}}
  \subfigure[\ $t = 0.33$]{\includegraphics[width=5cm]{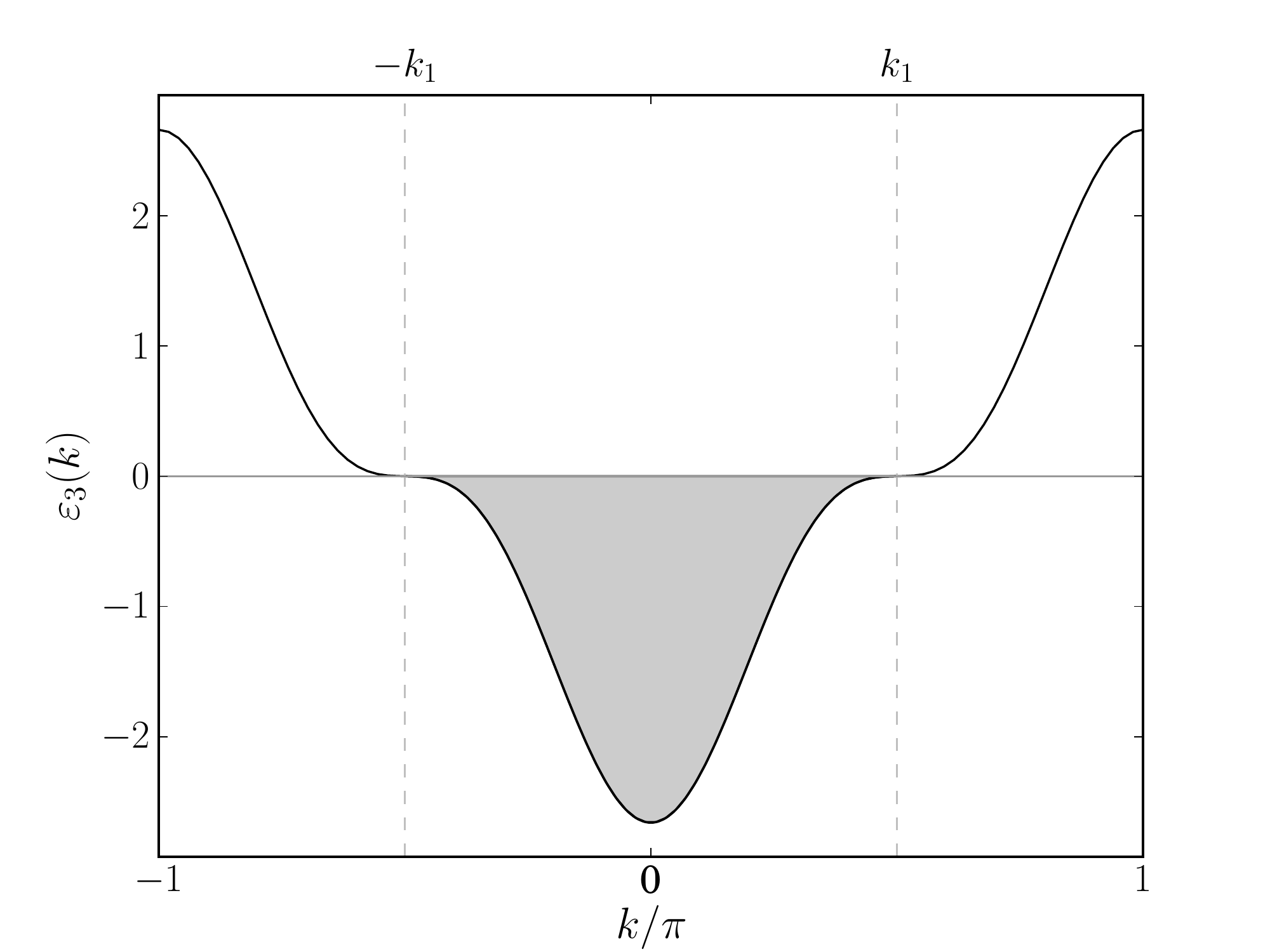}}
  \subfigure[\ $t = 0.6$]{\includegraphics[width=5cm]{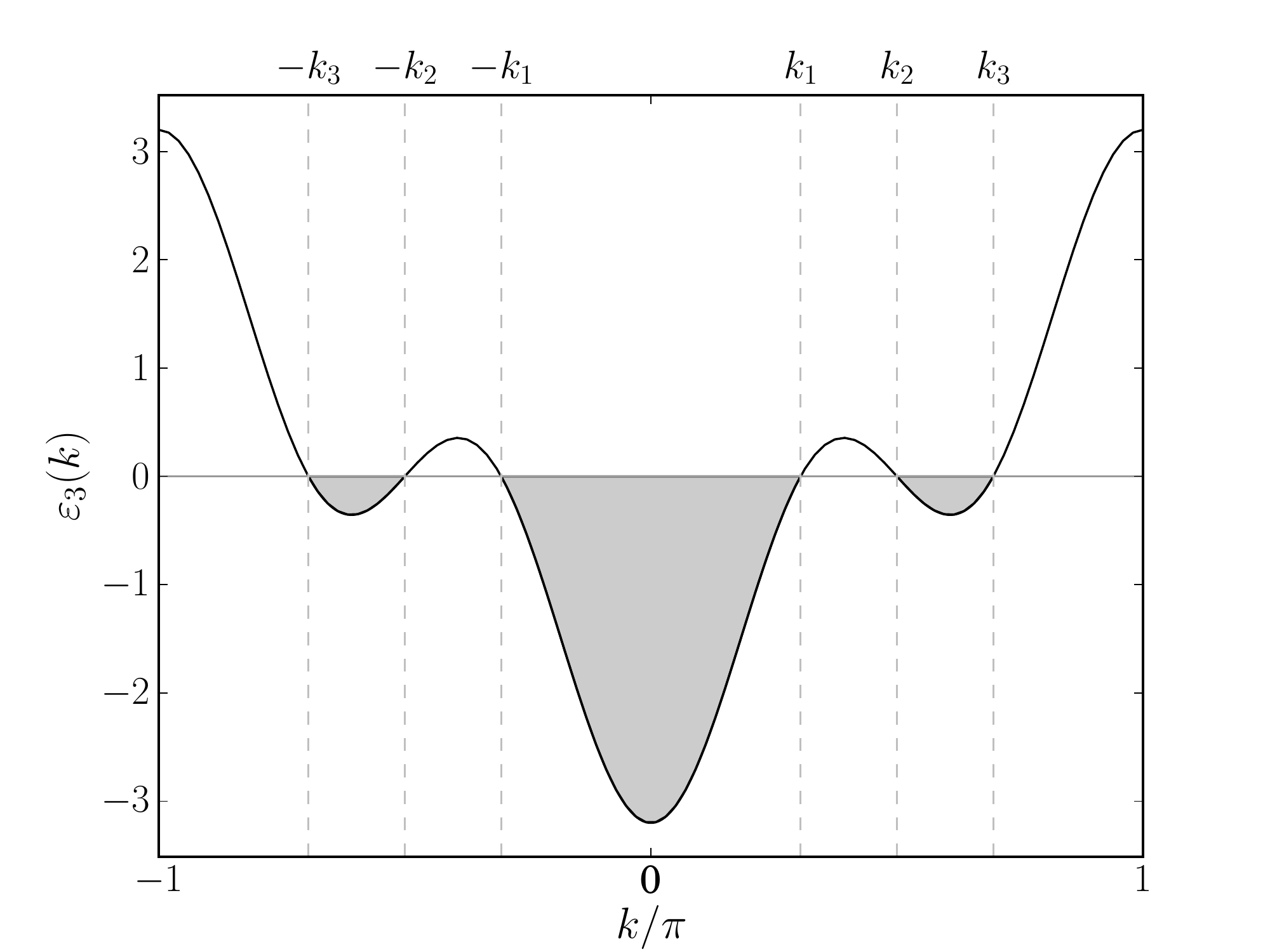}}
  \caption{(color online) Occupied modes (shaded) for the dispersion corresponding to Eq.~\eqref{eq:1d Hamiltonian} 
  with $n=3$.}
  \label{fig:fs1d-n3}
\end{figure*}

We begin with the simple case of one dimension. Consider the following, infinite system of one-dimensional fermions
\be
  \hat{H}_n = -\sum_i [(\hat{c}^\dag_i\hat{c}^{}_{i+1} + t\hat{c}^\dag_i\hat{c}^{}_{i+n}) + \text{h.c.}],
\label{eq:1d Hamiltonian}
\ee where $\hat{c}^\dag_i, \hat{c}^{}_i$ are fermionic creation and annihilation operators on site $i$ and $t$ is the amplitude of the $n$-th nearest-neighbor tunneling relative to the amplitude of nearest-neighbor tunneling. The system has the dispersion relation $\varepsilon_n(k) = -2[\cos k + t \cos (nk)]$. Since the dispersion is symmetric about $k=0$, let the occupied modes in the ground state be $k$ lying in the positive intervals $(0,k_1),\ (k_2,k_3), \ldots, (k_{m-1},k_m)$ if $m$ is odd and $(0,k_1),\ (k_2,k_3),\ldots (k_{m-2},k_{m-1}),\ (k_m,\pi)$ if $m$ is even, and similarly for negative values of $k$. Here we assume $t\geq0$. As shown for the case of $n=2$ in Fig.~\ref{fig:fs1d-n2} and $n=3$ in Fig.~\ref{fig:fs1d-n3}, $m$ is the number of Fermi surface ``components'' in the Brillouin zone, i.e., the Fermi surface has $2m$ points.

The entanglement entropy in the ground state of a system of non-interacting fermions can be determined from the matrix of Green's functions \cite{Peschel03}
\be
  M_{ij}(\Omega)
  = \mean{\hat{c}^\dag_i\hat{c}_j} 
  = \frac{1}{2\pi}\int_{-\pi}^\pi dk\ \Theta(-\varepsilon_n(k))e^{-ik(i-j)}
\ee for $i,j\in \Omega$, where $\Omega$ is the subsystem of interest and consists of an interval of length $\ell$ (lattice sites). Here $\Theta(x)$ is the Heaviside step function. The entanglement entropy is then given by
\be
  \mathcal{S}(\ell) = -\tr [M \ln M + (1-M)\ln(1-M)].
\label{eq:S of M}
\ee It is a remarkable fact that Eq.~\eqref{eq:S of M} can be evaluated analytically in the limit of large $\ell$, using techniques from the theory of Toeplitz, or translationally-invariant, matrices, in particular the Fisher-Hartwig conjecture \cite{Jin04,Keating}. A straightforward generalization of the calculation for the case $t=0$ carried out in Ref.~\cite{Jin04} reveals that for $t>0$ and $\ell\rightarrow\infty$ we have
\begin{widetext}
\be
  \mathcal{S}(\ell) = m\lp \frac{1}{3}\ln \ell + s_1 \rp 
  + \frac{1}{6}\lb \sum_{i=1}^m \ln (1-x_i^2) + 2\sum_{1\leq i < j \leq m} (-1)^{i+j} \ln \frac{1 - x_ix_j + (1-x_i^2)^{1/2}(1-x_j^2)^{1/2}}{1 - x_ix_j - (1-x_i^2)^{1/2}(1-x_j^2)^{1/2}} \rb,
\label{eq:ee 1D}
\ee
\end{widetext} where $s_1 = (\ln2)/3 + \Upsilon_1$ with
\be
  \Upsilon_1 = -\int_0^\infty dt\ \lb \frac{e^{-t}}{3t} + \frac{1}{t\sinh^2(t/2)} - \frac{\cosh(t/2)}{2\sinh^3(t/2)} \rb.
\ee 
The $x_j$'s are related to the Fermi momenta by $k_j=\cos^{-1} x_j$ where $|x_j|<1$ are the roots of $tT_n(x)+x=0$ with $T_n(x)$ the Chebyshev polynomials of the first kind.
Eq.~\eqref{eq:ee 1D} is valid if $\ell$ is sufficiently large relative to $|t-t_c|$, the distance from the critical point.
The main feature is the discontinuity of the logarithmic term: the prefactor of the leading, logarithmic term is proportional to the number of Fermi surface components $m$. 
The prefactor is proportional to the central charge of the underlying conformal field theory.
That this is the special case of a more general formula will become clear when we consider the Gioev-Klich formula in the next section.\\\\

\section{The Gioev-Klich formula in 1D and 2D}
Various arguments\cite{Wolf06,Gioev06,Li06,Swingle10,Cal12,Yang} have demonstrated that the leading scaling of the entanglement entropy in free fermions exhibits a logarithmic violation of the area law with subsystem size (linear dimension) $\ell$, $\mathcal{S}(\ell) \sim \ell^{d-1}\ln \ell$ in $d$ dimensions. More specifically, a conjecture by Widom \cite{Widom82} leads to the explicit formula \cite{Gioev06}
\begin{align}
  \mathcal{S}(\ell) &\sim \alpha \ell^{d-1}\ln \ell,\\
  \alpha &= \frac{1}{(2\pi)^{d-1}}\frac{1}{12}\int_{\d\Gamma} dA_k
  \int_{\d\Omega} dA_x\ |\hat{\vec{n}}_k \cdot \hat{\vec{n}}_x|,
\label{eq:widom}
\end{align}for the leading behavior of the entanglement entropy in an infinite system of gapless fermions with a Fermi surface of co-dimension 1, i.e., the surface separating the occupied and unoccupied regions of the $d$-dimensional Brillouin zone is a $(d-1)$-dimensional surface. Here $\d\Gamma$ is the Fermi surface, $\d\Omega$ is the real-space boundary of the subsystem $\Omega$ scaled such that the volume of $\Omega$ is 1, and $\hat{\vec{n}}_k$ and $\hat{\vec{n}}_x$ are unit vectors normal to $\d\Gamma$ and $\d\Omega$, respectively. Although the mathematical derivation of Eq.~\eqref{eq:widom} is somewhat complex, an intuitive explanation was given in Ref.~\cite{Swingle10} by associating each patch of the $(d-1)$-dimensional Fermi surface with a chiral conformal field theory that contributes $(1/6)\ln \ell$ to the total entanglement entropy, so that a simple patch-counting argument leads to the result.

In one dimension Eq.~\eqref{eq:widom} is simple to evaluate. For a real-space interval of length $\ell$ embedded in an infinite system, the geometric factor $\int_{\d\Gamma} dA_k\int_{\d\Omega} dA_x\ |\hat{\vec{n}}_k \cdot \hat{\vec{n}}_x|$ counts the number of boundary points on the Fermi surface and the real-space boundary, or $4m$ where $m$ is the number of Fermi surface components. Therefore
\be
  \alpha_\text{1D} = \frac{m}{3},
\ee which is consistent with Eq.~\eqref{eq:ee 1D} where the prefactor of the logarithmic scaling was derived through a microscopic calculation.

\begin{figure*}[t]
  \centering
  \subfigure[\ $t = -1.5$]{\includegraphics[width=4cm]{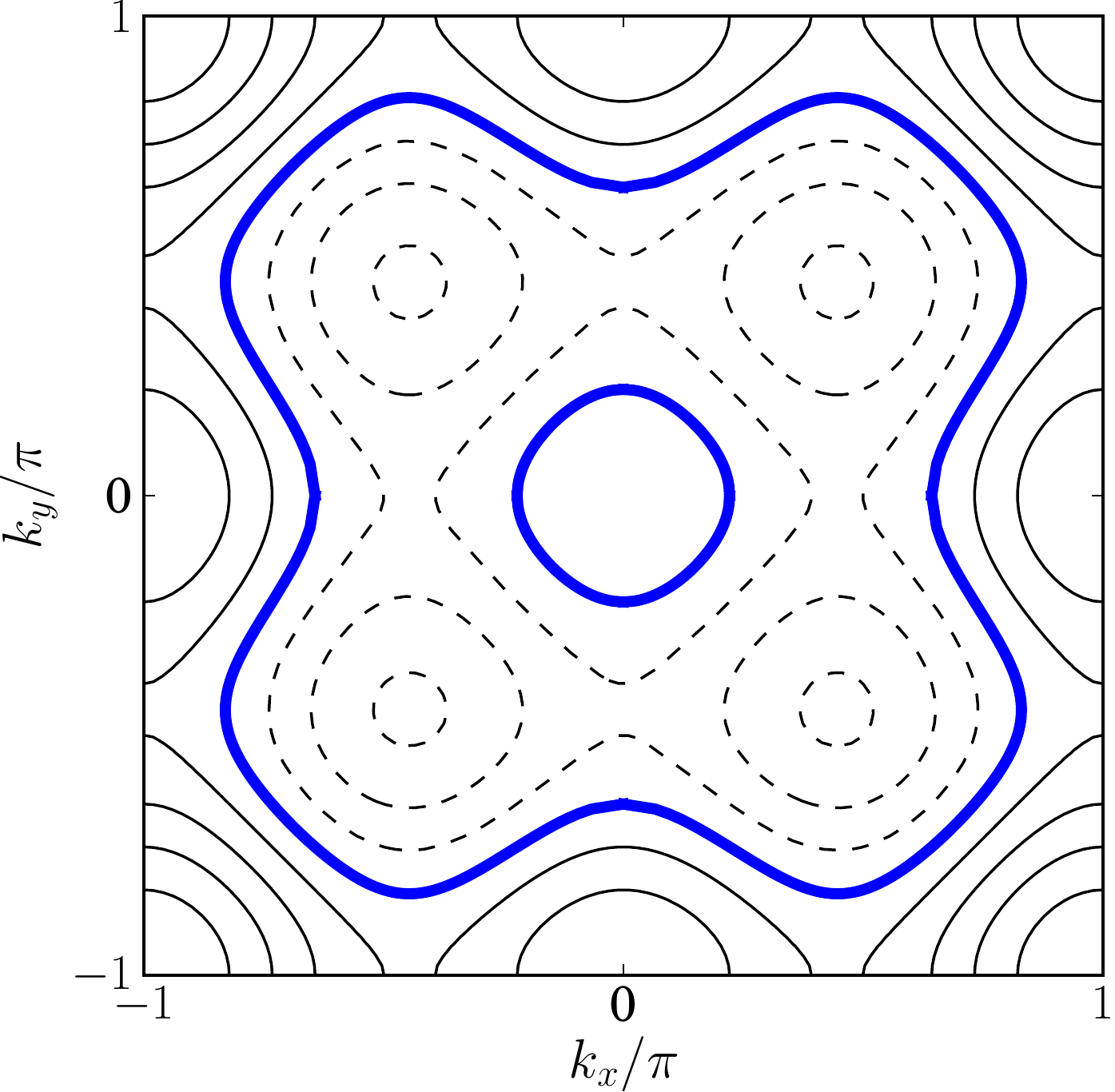}}
  \hspace{0.05\textwidth}\subfigure[\ $t = -0.5$]{\includegraphics[width=4cm]{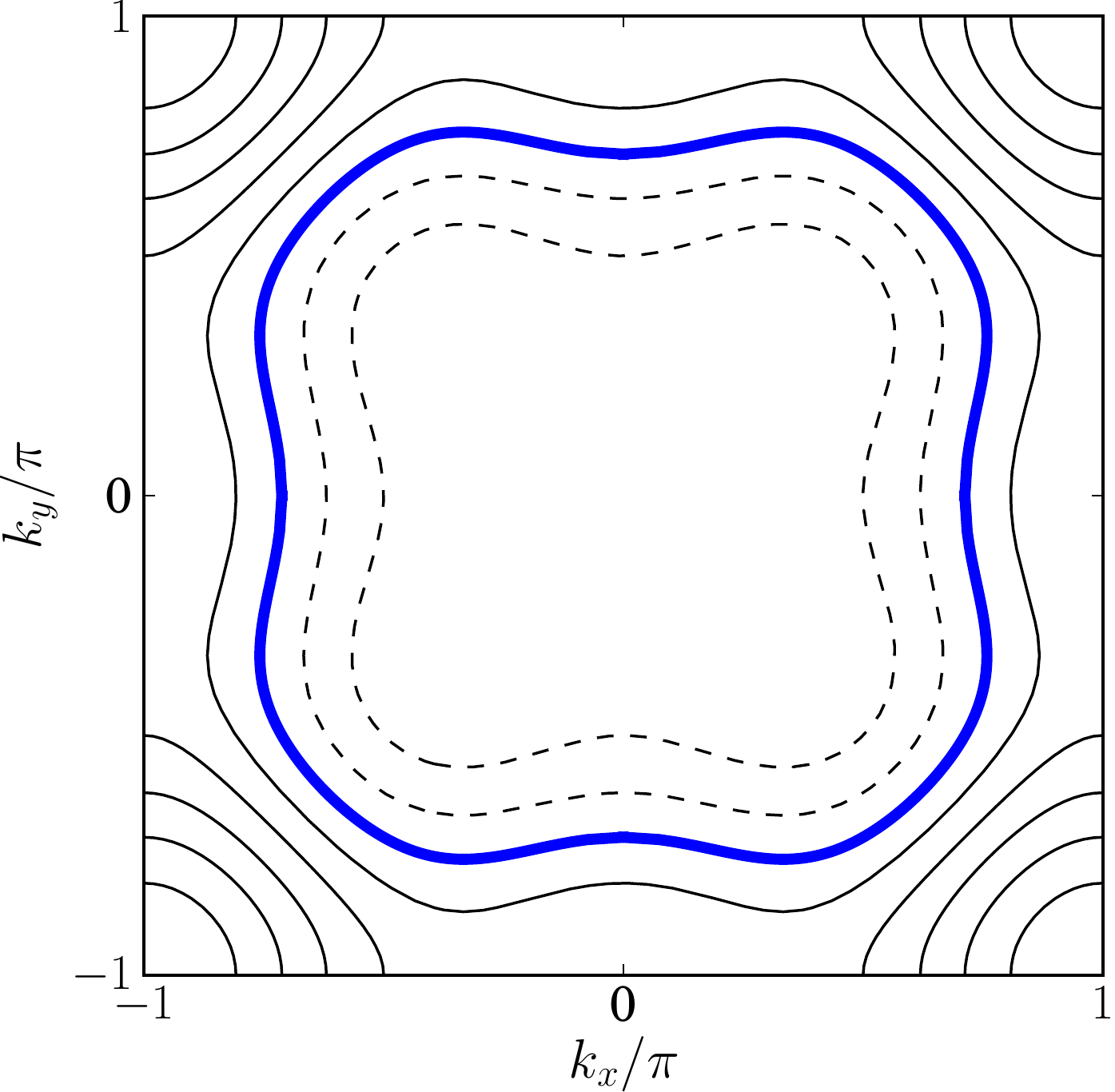}} 
  \hspace{0.05\textwidth}\subfigure[\ $t = -0.1$]{\includegraphics[width=4cm]{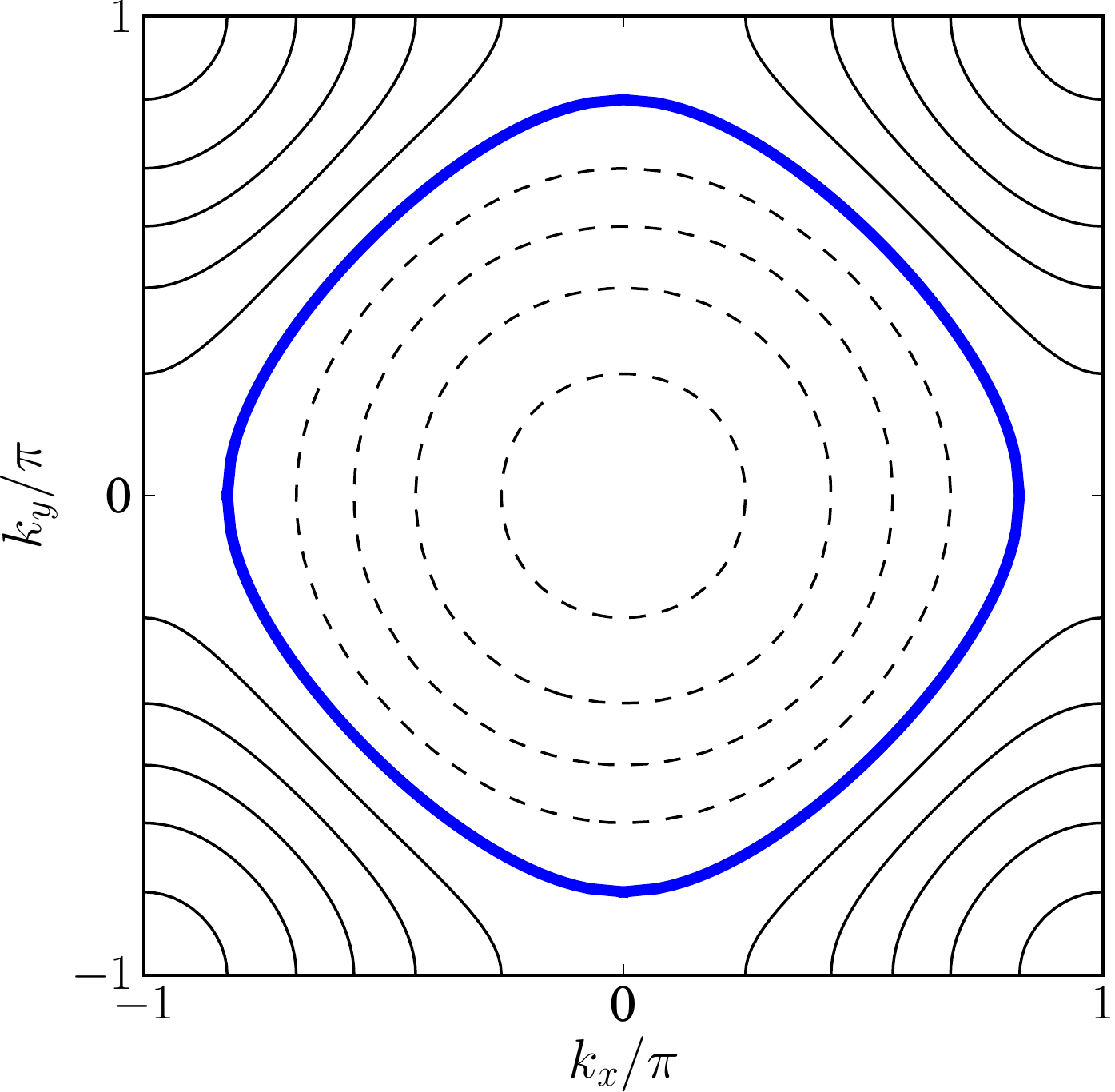}} \\
  \subfigure[\ $t = 0.1$]{\includegraphics[width=4cm]{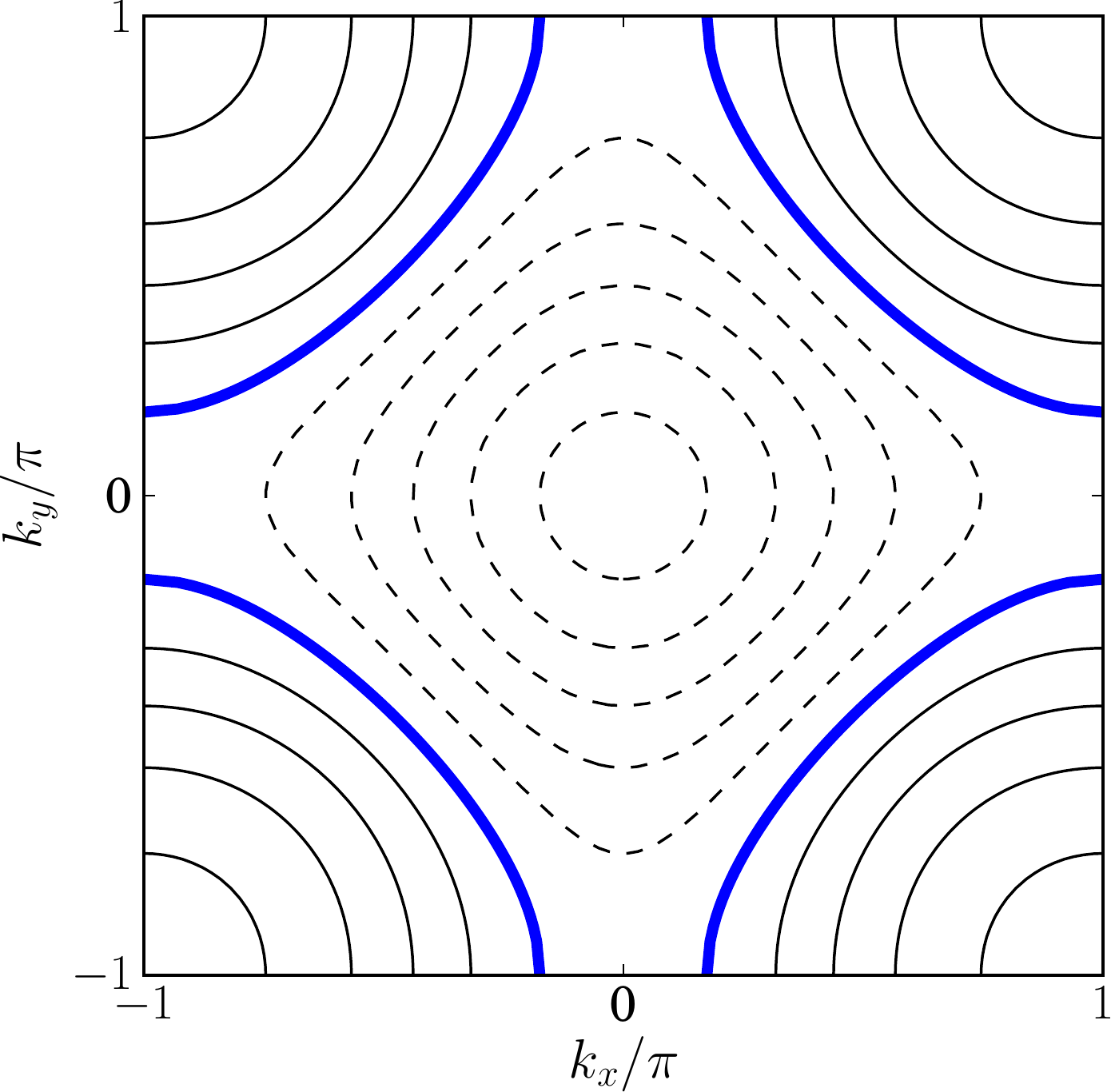}} 
  \hspace{0.05\textwidth}\subfigure[\ $t = 0.5$]{\includegraphics[width=4cm]{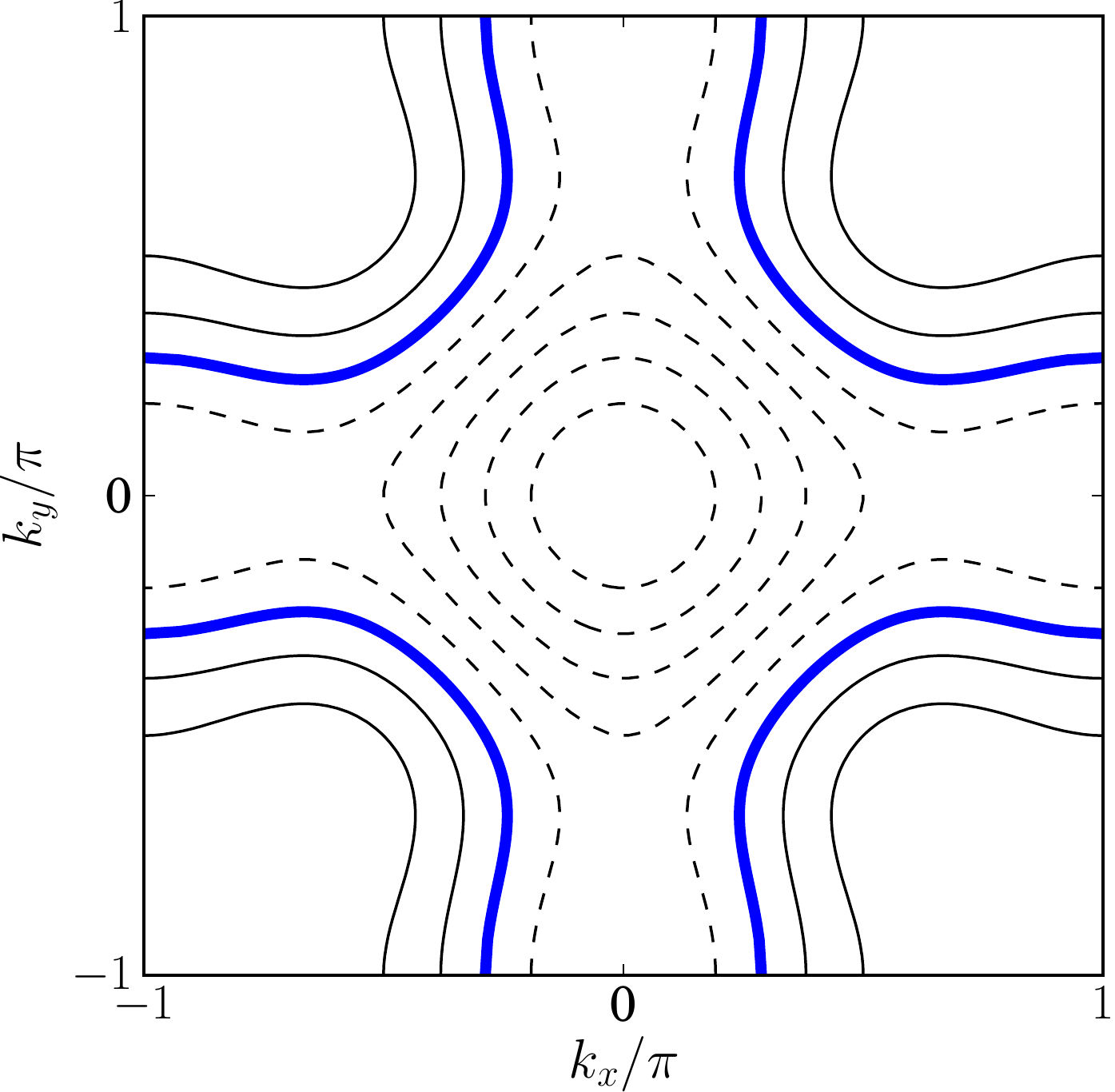}}
  \hspace{0.05\textwidth}\subfigure[\ $t = 1.5$]{\includegraphics[width=4cm]{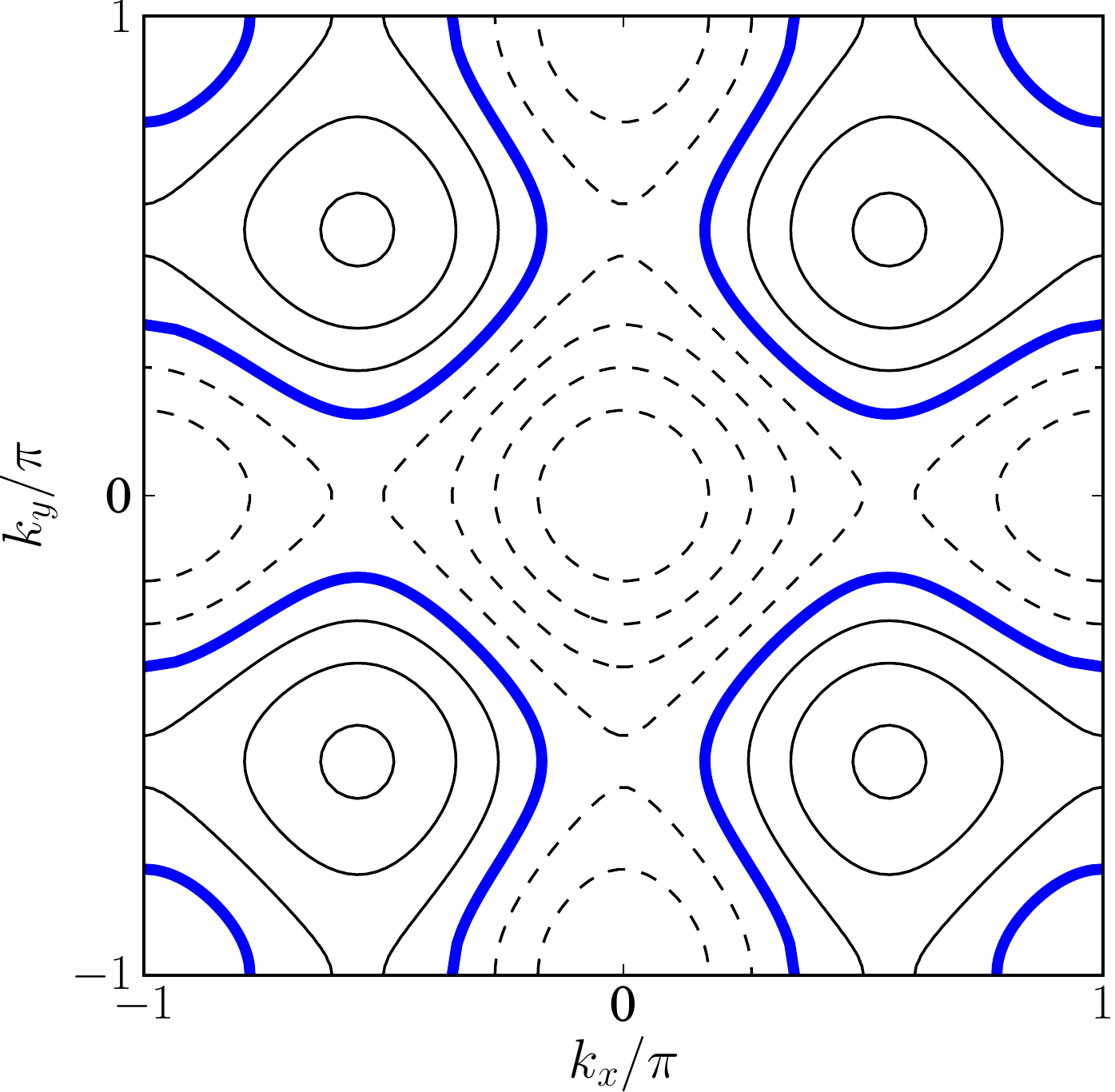}}
  \caption{(color online) Fermi surfaces (blue, thick lines) for the dispersion in Eq.~\eqref{eq:2d dispersion} with $\mu=0$.
 Solid (dashed) lines represent positive (negative) energy states.}
  \label{fig:fs2d}
\end{figure*}

In two dimensions Eq.~\eqref{eq:widom} can also be simplified. For convenience we consider an $\ell\times\ell$ square region of real space so that the scaled region $\Omega$ is a unit square. The normal unit vectors in real space are thus $\pm\vec{x},\pm\vec{y}$. Since the Fermi surface is a curve in two dimensions, we can parametrize the curve by $k_x=k_x(\theta)$ and $k_y=k_y(\theta)$, with the unit normal vector given by the normalized velocity $\hat{\vec{n}}_k=\nabla_\vk\varepsilon/|\nabla_\vk\varepsilon|$ where $\varepsilon(\vk)$ is the dispersion relation. Then
\be
  \int_{\d\Omega} dA_x\ |\vh{n}_k \cdot \vh{n}_x|
  = 2\frac{1 + \big|\pder{\varepsilon}{k_y}\big/\pder{\varepsilon}{k_x}|}{\sqrt{1 + \big(\pder{\varepsilon}{k_y}/\pder{\varepsilon}{k_x}\big)^2}}.
\ee Meanwhile, the line element is given by
\be
  dS_k = \sqrt{\lp\der{k_x}{\theta}\rp^2 + \lp\der{k_y}{\theta}\rp^2}\ d\theta,
\ee but since
\be
  \pder{\varepsilon}{\theta} = \pder{\varepsilon}{k_x}\der{k_x}{\theta} + \pder{\varepsilon}{k_y}\der{k_y}{\theta} = 0
\ee along the Fermi surface, we have
\be
  \frac{\d\varepsilon/\d k_y}{\d\varepsilon/\d k_x}
  = -\frac{dk_x/d\theta}{dk_y/d\theta}
\ee and we can write
\be
  \alpha_\text{2D} = \frac{1}{12\pi}\int_{\d\Gamma} d\theta\ \lp \bigg|\der{k_x}{\theta}\bigg| + \bigg|\der{k_y}{\theta}\bigg| \rp.
\label{eq:alpha 2d}
\ee Eq.~\eqref{eq:alpha 2d} emphasizes the purely geometrical character of the prefactor of the leading scaling of the entanglement entropy in such systems, and can be evaluated by splitting the parameterization of the Fermi surface into piece-wise monotonic regions. 
As the simplest example, 
consider the case of nearest-neighbor hopping on the square lattice at zero chemical potential. 
Then the dispersion is $\varepsilon(\vk)=-2(\cos k_x + \cos k_y)$. 
Since the Fermi surface is invariant under 90-degree rotations (which we will assume in all of the models considered in this work), we can treat the first quadrant only and multiply by 4. 
In the first quadrant of the Brillouin zone the Fermi surface is clearly parameterized by $k_x = \theta,\ k_y = \pi - \theta$ for $\theta\in [0,\pi)$, so that Eq.~\eqref{eq:alpha 2d} gives
\be
  \alpha_\text{2D} = \frac{1}{3\pi}\lb k_x(\theta)\bigg|_0^\pi - k_y(\theta)\bigg|_0^\pi \rb = \frac{2}{3}.
\ee More generally, including a chemical potential term in the dispersion leads to the prefactor \cite{Li06,Barthel06}
\be
  \alpha(\mu) = \frac{2}{3}\lb 1 - \frac{1}{\pi}\cos^{-1}\lp 1 - \frac{|\mu|}{2} \rp \rb
\ee for $|\mu|\leq 4$, which has a cusp at $\mu=0$ such that near $\mu=0$ we have $\alpha(\mu) \approx \alpha(0) - (2/3\pi)|\mu|^{1/2}$. Thus at $\mu=0$ we have a simple example of a Lifshitz transition, and in fact this transition is very similar in character to the transition at $t=0$ studied below, cf. Figs.~\ref{fig:fs2d}(c),\ref{fig:fs2d}(d).


Consider now the more interesting dispersion relation
\begin{align}
  \varepsilon(\vk) &= -2(\cos k_x + \cos k_y) \notag\\
  &\qquad - 2t [\cos (2k_x) + \cos (2k_y)] - \mu,
\label{eq:2d dispersion}
\end{align} which arises from nearest-neighbor and next-\emph{next}-nearest neighbor coupling on the square lattice. We treat this case to simplify the algebra, although a slightly more realistic model would include next-nearest neighbor coupling instead. To parametrize the Fermi surface, we again use the 90-degree rotational symmetry to limit ourselves to the first quadrant of the Brillouin zone and multiply the overall result by 4. Let (we assume $t\neq0$ for the moment)
\begin{align}
  C(t) &= -\frac{1}{4t},\\
  R(t,\mu) &= \sqrt{1 + \mu C(t) + 2C(t)^2}.
\end{align} Then the Fermi surface is defined by
\begin{align}
  k_x(\theta) &= \cos^{-1}[C(t) + R(t,\mu)\cos\theta],\\
  k_y(\theta) &= \cos^{-1}[C(t) + R(t,\mu)\sin\theta]
\end{align} with the ranges of $\theta$ depending on $t,\mu$ so that $-1 \leq C + R\cos\theta \leq 1$ and $-1 \leq C +R\sin\theta \leq 1$. For example, if $\mu=0$ then
\be
  \theta \in \begin{cases}
    (\pi/2 + W(t),\ 2\pi - W(t)) \text{ and}\\
    \qquad (W(t),\ \pi/2 - W(t)) & \text{ for $t<-1$},\\
    (\pi/2 + W(t),\ 2\pi - W(t)) & \text{ for $-1\leq t < 0$},\\
    (-\pi/2 + W(t),\ \pi - W(t)) & \text{ for $0 < t \leq 1$},\\
    (-\pi/2 + W(t),\ \pi - W(t)) \text{ and}\\
    \qquad (-\pi + W(t),\ -\pi/2 - W(t)) & \text{ for $t > 1$},
  \end{cases}
\ee where
\be
  W(t) = \cos^{-1} \frac{1 + C(|t|)}{R(|t|)}.
\ee We note that the ranges for $t>0$ are merely shifted by $\pi$ from the values for $t<0$. Fig.~\ref{fig:fs2d} shows some representative Fermi surfaces for different values of $t$. For $t<-1$ the occupied modes contain a ``hole'' that disappears at $t=-1$, while at $t=1$ disconnected Fermi components appear.


\begin{figure}[h]
  \includegraphics[width=7cm]{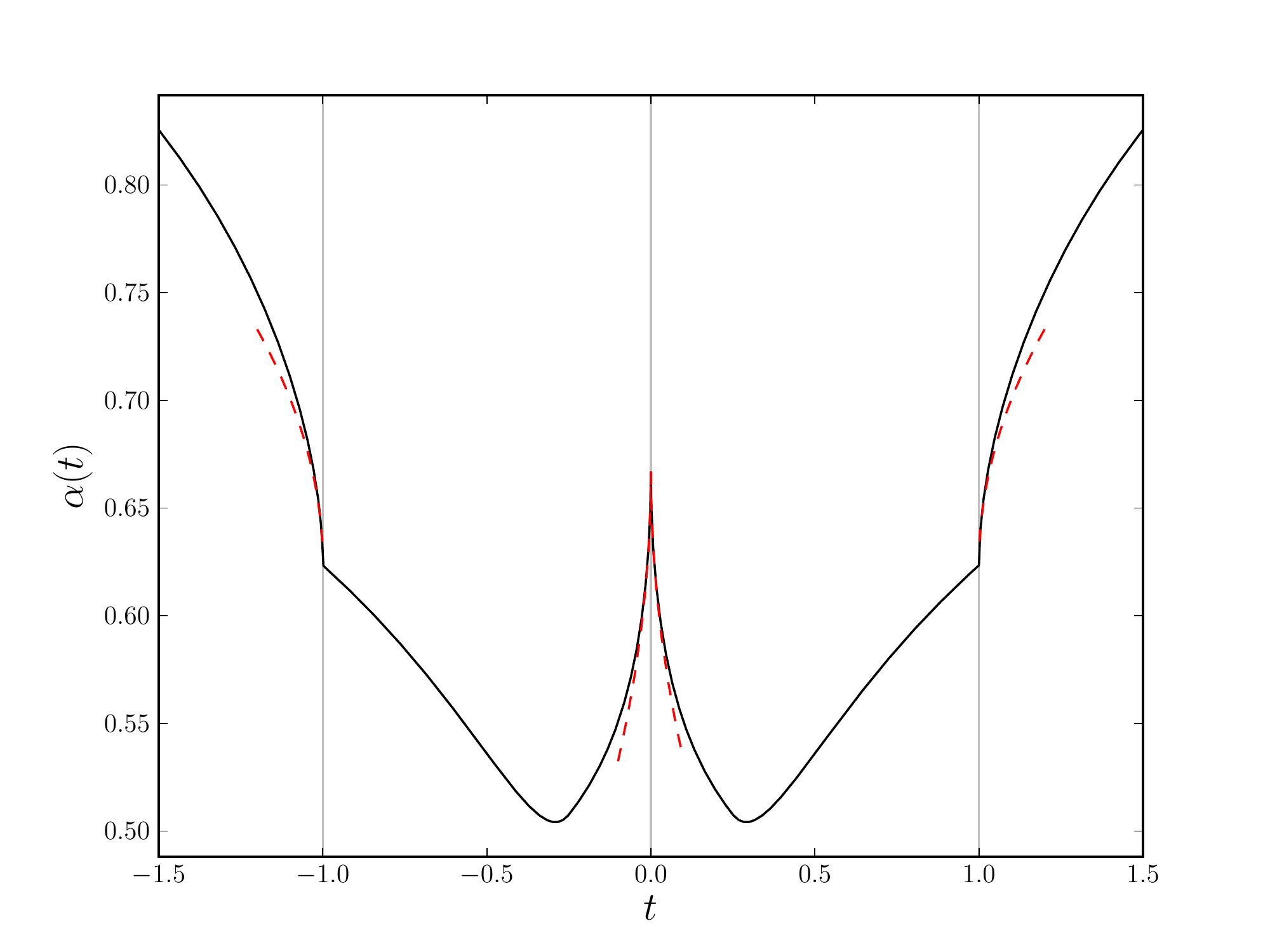}
  \caption{The coefficient of the leading contribution to the entanglement entropy, given by Eq.~\eqref{eq:alpha t 2d}. Dashed lines show the square-root nature of the kinks/cusps.}
  \label{fig:alpha 2d}
\end{figure}

The result of using Eq.~\eqref{eq:alpha 2d} is invariant under $t\rightarrow -t$ and can be described as
\be
  \alpha(t) = \frac{2}{3} \times \begin{cases}
    1 - f(W(t),t) ~~~~~~~~~~~~~~ \text{for $0 < |t| < 1/4$},\\
    1 - f(W(t),t) + 2f(W(t),t)- 2f(\pi/2,t) \\
~~~~~~~~~~~~~~~~~~~~~~~~~~~~~~~~  \text{for $1/4 \leq |t| \leq 1$},\\
    1 - f(W(t),t) + 2f(W(t),t)- 2f(\pi/2,t) \\
 - f(W(t) - \pi,t) + 1 ~~~~~~ \text{for $|t| > 1$},
  \end{cases}
\label{eq:alpha t 2d}
\ee
where
\be
  f(\theta, t) = \frac{1}{\pi}\cos^{-1}[ C(|t|) + R(|t|)\sin\theta ].
\ee As shown in Fig.~\ref{fig:alpha 2d} the prefactor of the logarithmic scaling is continuous but non-monotonic even for the same sign of $t$ (partly because particle number is not fixed as $t$ changes), 
and moreover has square-root singularities at $t=0$ and $|t|=1$. Specifically, near $t=0$ we have $\alpha(t)\approx\alpha(0) -(4/3\pi)|t|^
{1/2}$ while near $|t|\gtrsim 1$ we have $\alpha(t)\approx \alpha(1) + (4/3\sqrt{3}\pi)(|t|-1)^{1/2}$. The presence of square-root kinks is a general feature in 2D, and is explained in the following section for arbitrary dimensions.

\section{General dimensions}
We consider here the scaling of entanglement entropy near a Lifshitz transition in dimension $d$, focusing on the case where the co-dimension of the Fermi surface is 1 and the dynamical critical exponent at the critical point is 2, but it is possible and straightforward to extend the result further to more general cases. Consider the local fermion dispersion
\be
\varepsilon(k) = \sum_{i=1}^p k_i^2  - \sum_{j=p+1}^d k_j^2 + \delta t,
\ee
where $k_i$ is the $i$-th component of momentum measured from the momentum point
where the Lifshitz transition occurs at $\delta t=0$ and $p$ is an integer that parameterizes the topology of the Fermi surfaces involved in the transition.
Locally in momentum space, the dispersion respects SO$(p,d-p)$ symmetry.
For $p=0$ ($p=d$), this dispersion describes the transition where
a new hole-like (electron-like) Fermi surface of topology $S^{d-1}$ 
appears as $\delta t$ becomes positive (negative).
For $0< p < d$, it describes the transition where
the neck of the hyperbolic Fermi surface pinches to a point to change its topology.
Near the critical point, the area of the Fermi surface 
in momentum space changes non-analytically as 
\be
A \sim | \delta t |^{(d-1)/2} \Theta( \pm \delta t)
\ee
for $p=0$ or $d$, and
\be
A \sim | \delta t |^{(d-1)/2}
\ee
for $0< p < d$.
In $d=1$, it becomes a jump as we saw in Sec. II.
It is of note that the non-analyticity does not depend on $p$.
Since the entanglement entropy is related to the area of the Fermi surface (specifically, each patch of Fermi surface contributes a logarithm \cite{Swingle10}), the same non-analyticity shows up in the scaling of entanglement entropy.

\section{Numerical Results}

\subsection{1D}

\begin{figure}[h]
  \centering
  \subfigure[]{\includegraphics[width=5cm, height=5cm]{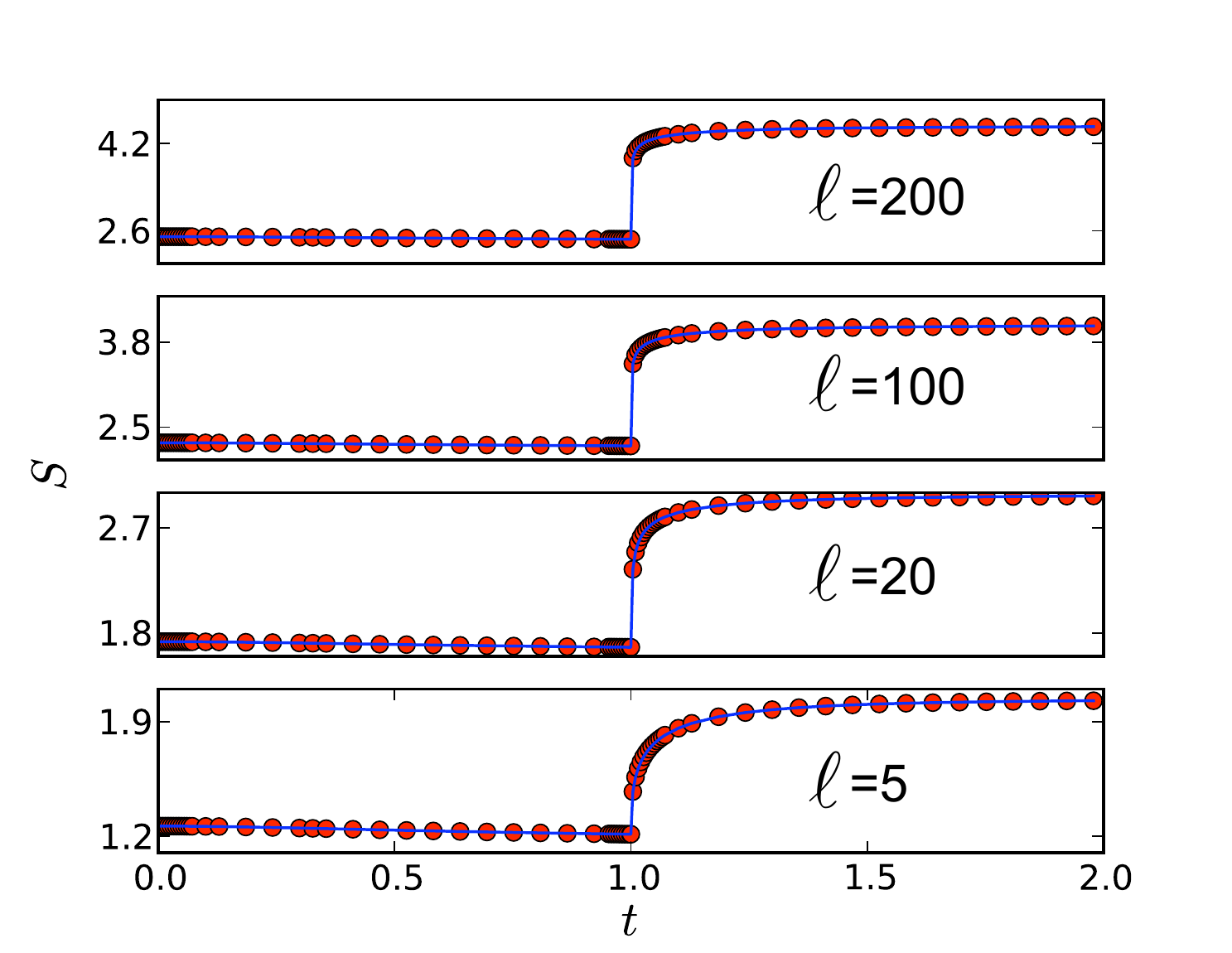}}
  \subfigure[]{\includegraphics[width=5cm, height=5cm]{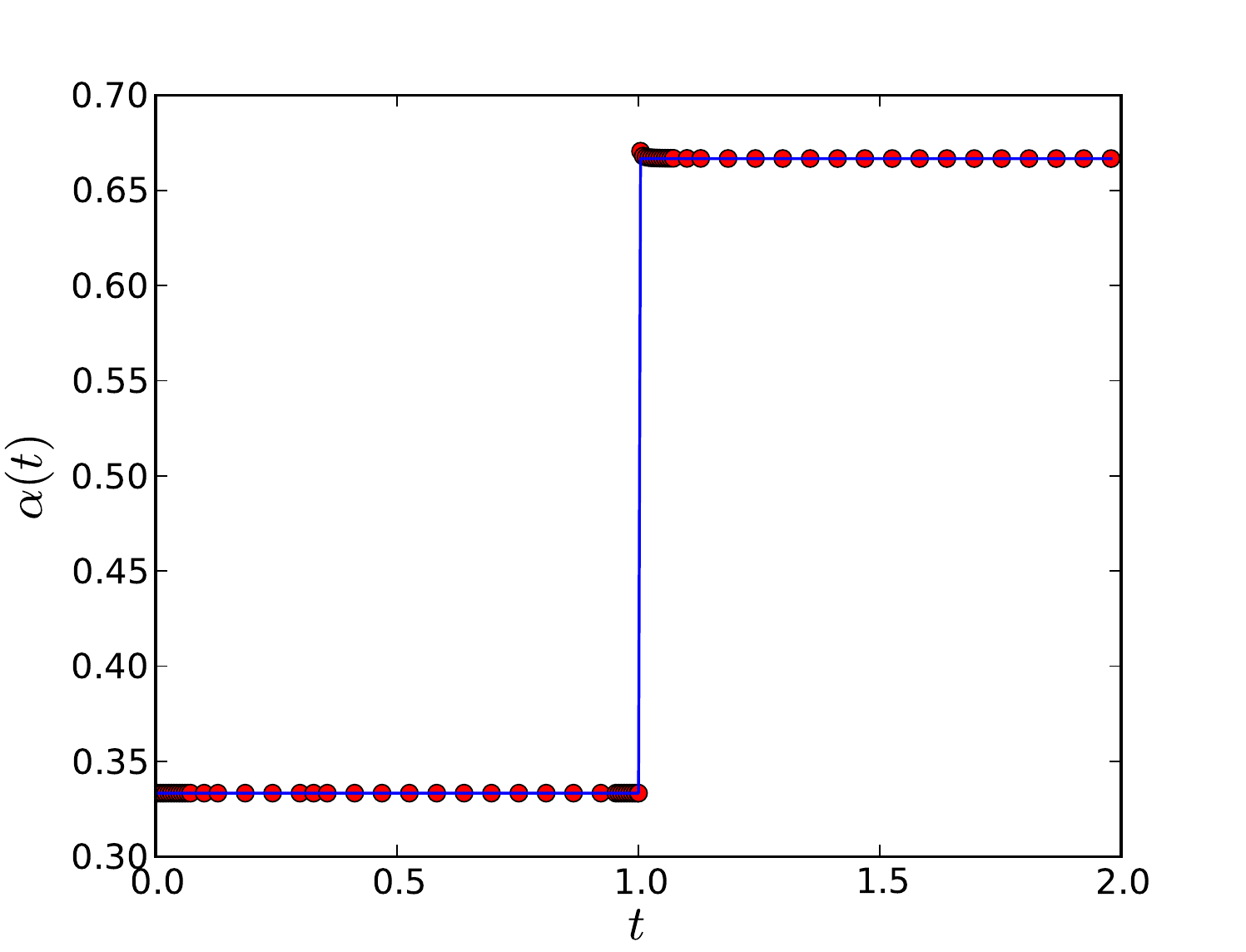}}
  \subfigure[]{\includegraphics[width=5cm, height=5cm]{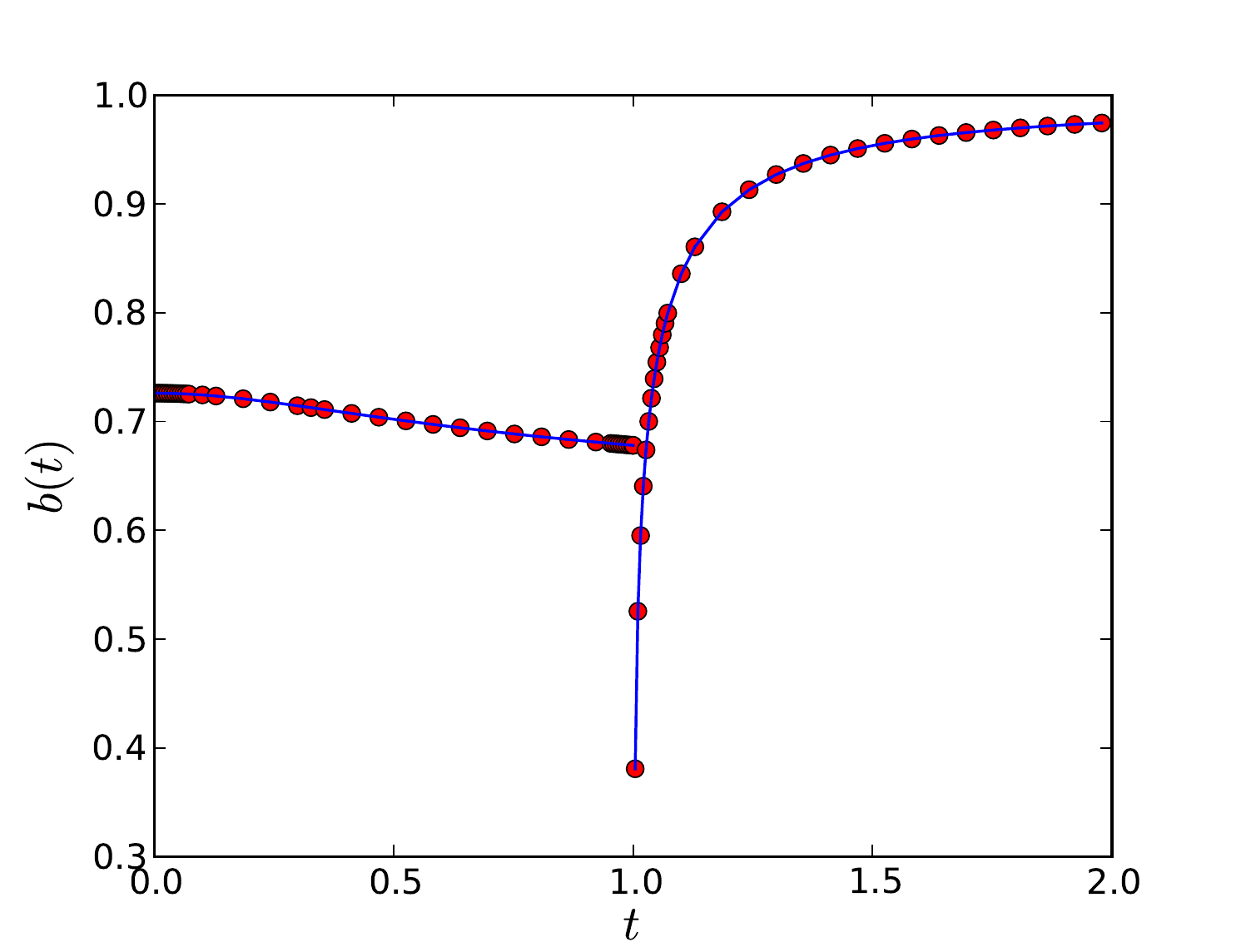}}
  \caption{(Color online) 
(a) Numerically calculated entanglement entropy in one dimension as a function of 
the next-nearest neighbor hopping $t$ ($n=2$) for various values of subsystem size. 
(b) The leading coefficient, $\alpha(t)$ as a function of $t$.  
(c) The sub-leading coefficient, $b(t)$ as a function of $t$. 
The solid lines are guides to the eye. 
}
  \label{fig:1D2}
\end{figure}

\begin{figure}[h]
  \centering
  \subfigure[]{\includegraphics[width=5cm, height=5cm]{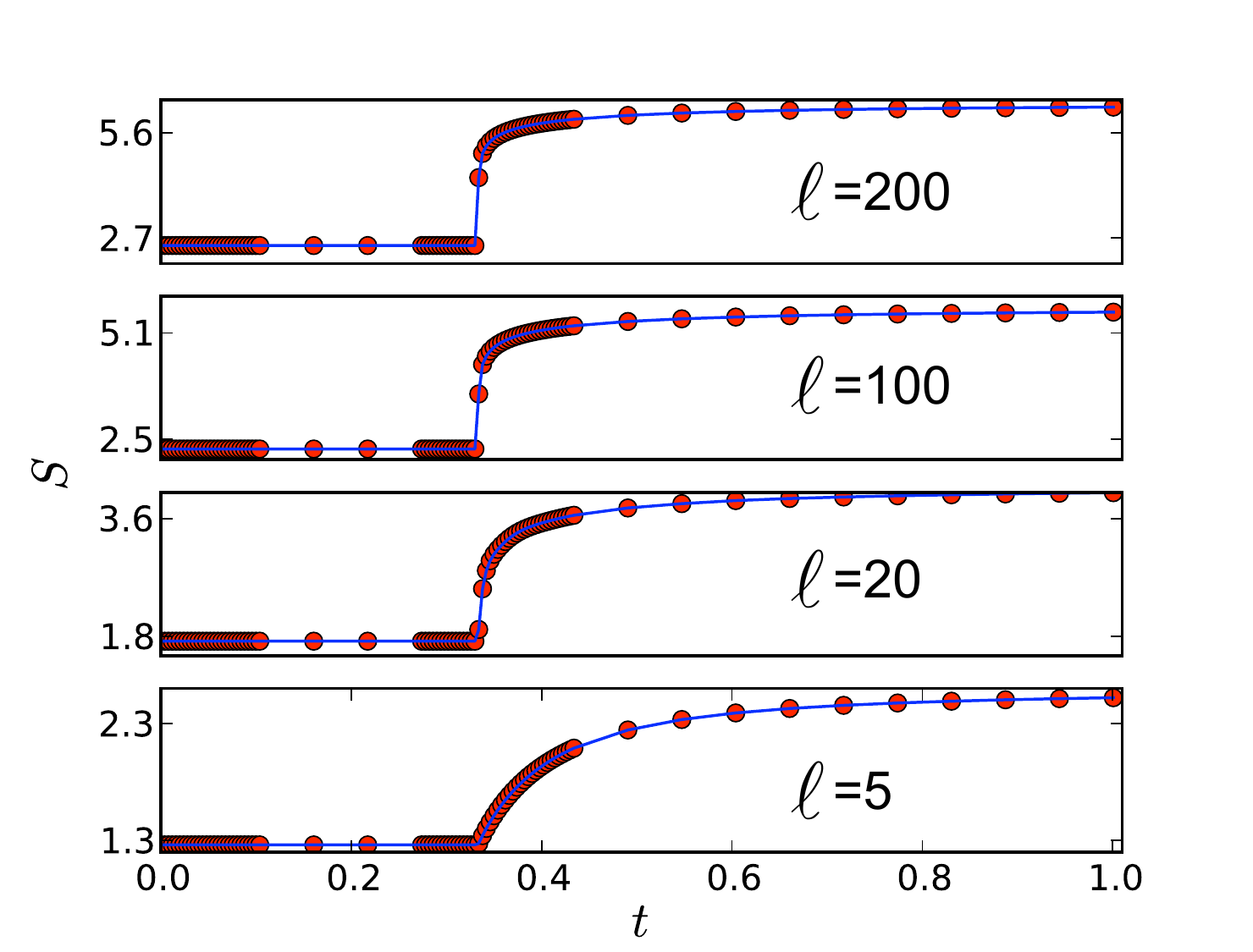}}
  \subfigure[]{\includegraphics[width=5cm, height=5cm]{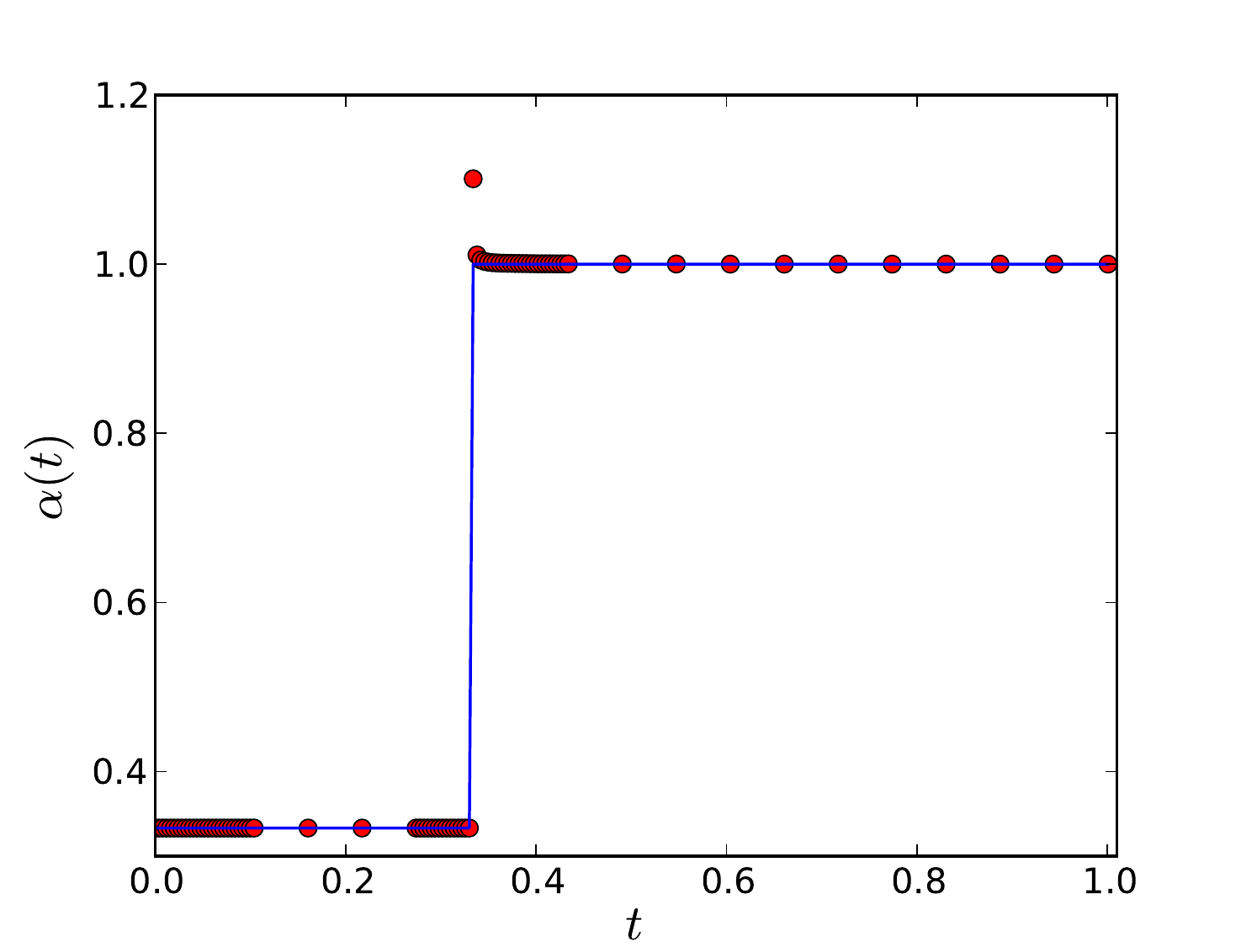}}
  \subfigure[]{\includegraphics[width=5cm, height=5cm]{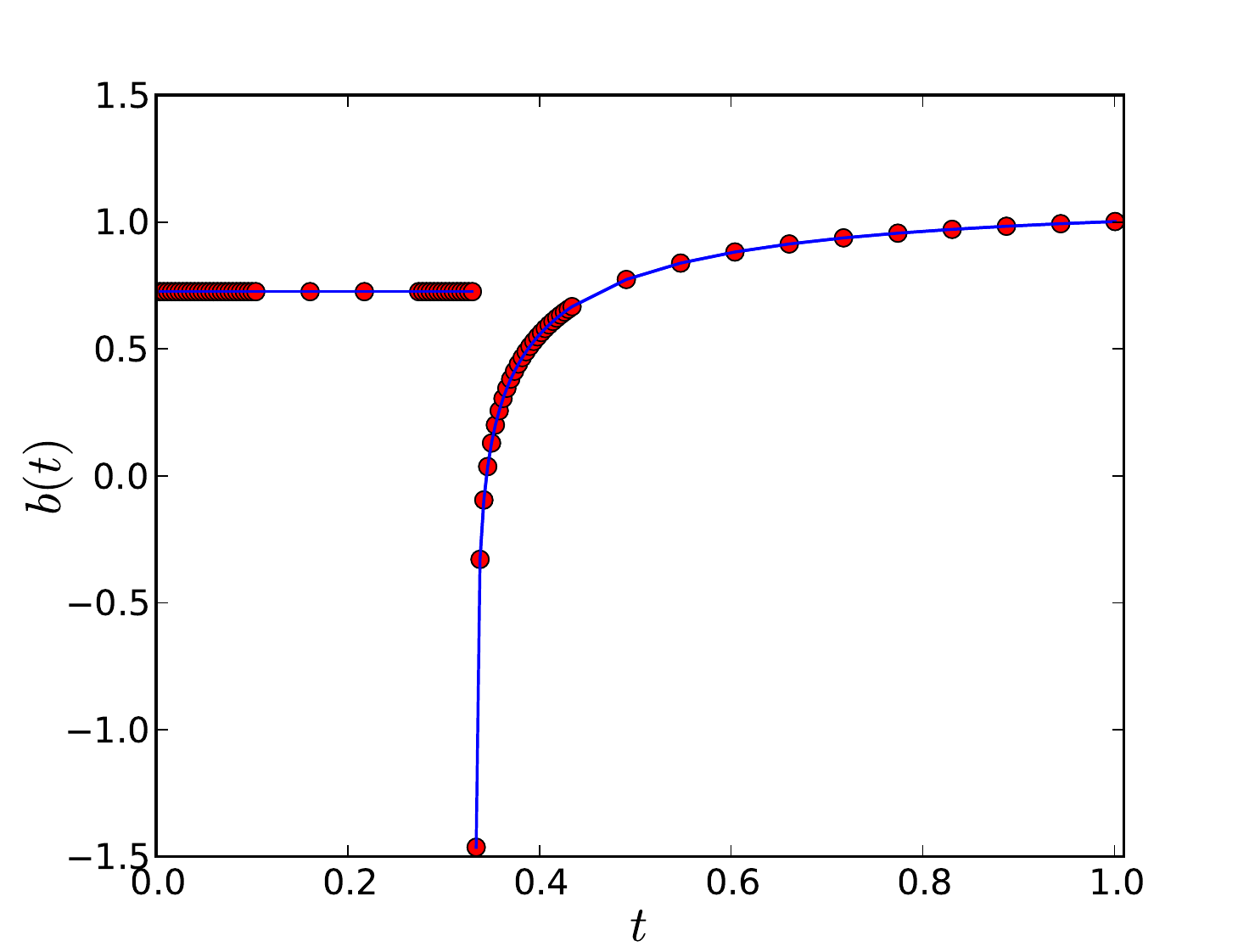}}
  \caption{(Color online) (a) Numerically calculated entanglement entropy in one dimension as a function of 
the next-next nearest neighbor hopping $t$ ($n=3$) for various values of subsystem size. 
(b) The leading coefficient, $\alpha(t)$, as a function of $t$.  
(c) The sub-leading coefficient, $b(t)$, as a function of $t$. 
The solid lines are guides to the eye. 
}
  \label{fig:1D3}
\end{figure}

In this section we present results for the entanglement entropy computed numerically for the one-dimensional Hamiltonian in Eq.~\eqref{eq:1d Hamiltonian}. Numerical data is obtained for system sizes ranging from $\ell$ = 20 to $\ell$ = 200 for $n$ = 2,3 respectively.
In Fig.~\ref{fig:1D2}(a), we show the entanglement entropy for $n=2$ as a function of $t$ for various values of $\ell$.
Even for relatively small $\ell$, the Lifshitz transition  at $t=1$ where the number of Fermi points changes is clearly visible.
As $\ell$ increases, the discontinuity becomes more pronounced and sharper.
The shape of the entanglement entropy is pretty similar for $\ell=100$ and $200$,
which suggests that $\ell=100$ is already close to the thermodynamic limit.
The entanglement entropy as a function of $\ell$ is fit to the form 
$\mathcal{S}(\ell) = \alpha(t) \ln\ell + b(t)$ upto $\ell=200$, 
allowing us to extract the coefficients as a function of $t$.
It is noted that $\alpha(t)$ and $b(t)$ 
are the coefficients obtained by fitting $\mathcal{S}(\ell)$
as a function of $\ell$.
Therefore, they are independent of $\ell$.
The coefficients are shown in Fig.~\ref{fig:1D2}(b) and Fig.~\ref{fig:1D2}(c).
As expected from the increase of the number of Fermi points by a factor of $2$, 
the leading coefficient $\alpha(t)$ doubles across the transition.
In Figs.~\ref{fig:1D3}(a), \ref{fig:1D3}(b), and \ref{fig:1D3}(c), 
we show the entanglement entropy and the leading coefficients in the $n$ = 3 case. 
Because the number of Fermi points increase by factor of $3$,
$\alpha(t)$ triples in magnitude across the transition.

\begin{figure}
  \includegraphics[width=7cm, height=7cm]{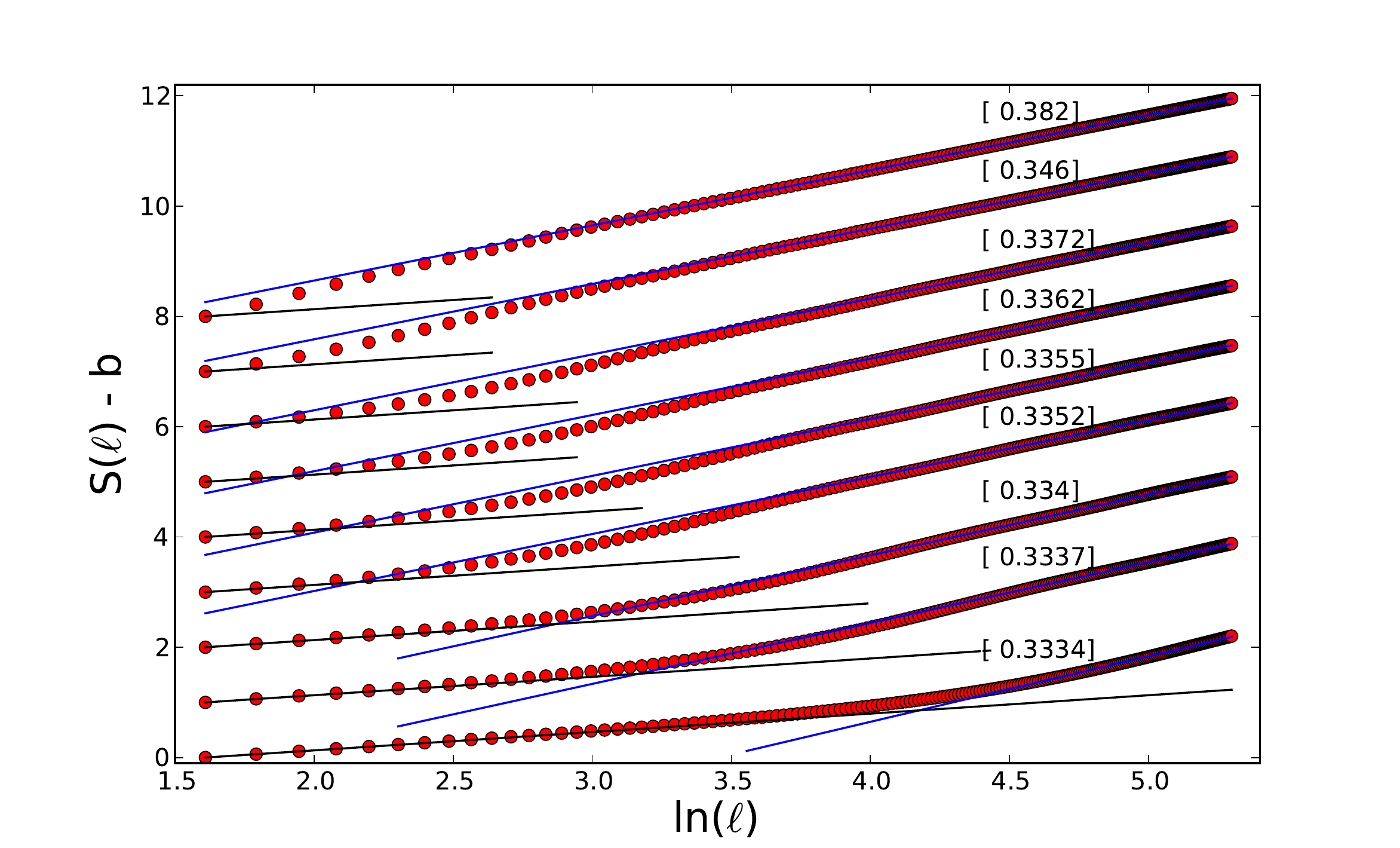}
  \caption{(color online) 
The entanglement entropy with the subtraction of the constant piece plotted as a function of $\ln \ell$ 
near the critical point $t_c=1/3$ with $n=3$.  
The number above each curve denotes the value of $t$.
The entanglement entropy is shifted vertically by a different amount for each $t$
to differentiate different curves.
Here $\ell$ ranges from $5$ to $200$.
For large $\ell$, the entanglement entropy fits well with $\ln \ell$ (blue line),
whereas for small $\ell$, the entanglement entropy behaves as $\frac{1}{3} \ln \ell$ (black line).
}
  \label{fig:LS0}
\end{figure}

Now we discuss the finite size 
crossover behavior of entanglement entropy 
near the critical point focusing on the case with $n=3$.
When $t$ is slightly larger than the critical hopping $t_c=1/3$, 
two small Fermi pockets are formed near $k=\pm \pi/2$ as is 
shown in Fig. 2. 
This sets a characteristic length scale 
$\ell^* \sim k_F^{-1} \sim (t-t_c)^{-1/2}$,
where $k_F$ is the size of the newly formed pockets. 
This can be understood from the 
dispersion $\epsilon_k = a( t_c - t ) \delta k + b\ \delta k^3 + ...$,
with $\delta k = k - \pi/2$ near the critical point,
which implies that the size of small Fermi pockets
scale as $k_F \sim ( t - t_c )^{1/2}$.
For $l > l^*$, the entanglement entropy is expected to show the 
asymptotic behavior $\frac{c}{3} \ln \ell$,
where $c=3$ is the central charge for the low energy modes.
For $\ell < \ell^*$, however, the newly formed small Fermi surfaces
are not completely resolved, and the entanglement entropy
shows a different behavior.
In Fig. \ref{fig:LS0}, we plot the entanglement entropy
with the constant piece subtracted, as a  function of 
$\ln \ell$ for various values of $t$ with $n=3$.
When $t$ is sufficiently close to $t_c$, 
there is a wide range of $\ell < \ell^*$ where 
the small Fermi pockets are invisible so 
that the entanglement entropy increases
as $\frac{1}{3} \ln \ell$
before it crossovers to $\ln \ell$.
For $t > 0.336$, the $\frac{1}{3} \ln \ell$ behavior
essentially disappears.
Interestingly, the entanglement entropy
appears to increase with a higher power of the logarithm 
$( \ln \ell )^A$ with $A>1$ at short distances
before it saturates to $\ln \ell$. 
In the non-interacting case with $n=3$, 
the ground state right at the critical point is rather boring:
it is the same as the one for $t < t_c$.
That is why the entanglement entropy at the critical point
shows the usual logarithmic behavior with $c=\frac{1}{3}$.
However, this is no longer the case for interacting systems.
It would be of great interest in the future to study
the possibility of the violation of the logarithmic behavior
in interacting quantum field theories with 
general dynamical critical exponent $z > 1$.

\subsection{2D}

%

\begin{figure}[h]
  \centering
  \subfigure[]{\includegraphics[width=5cm, height=5cm]{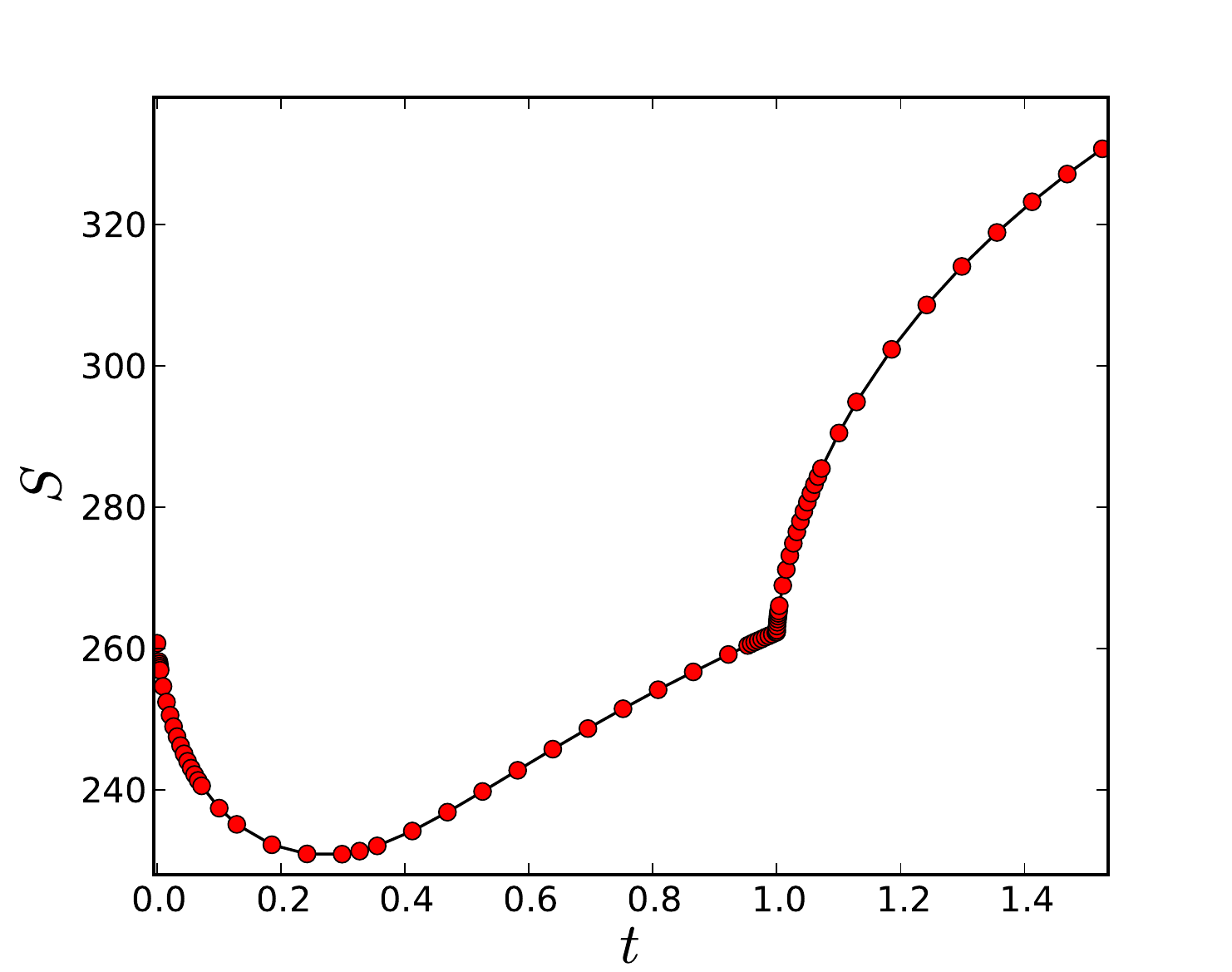}}
  \subfigure[]{\includegraphics[width=5cm, height=5cm]{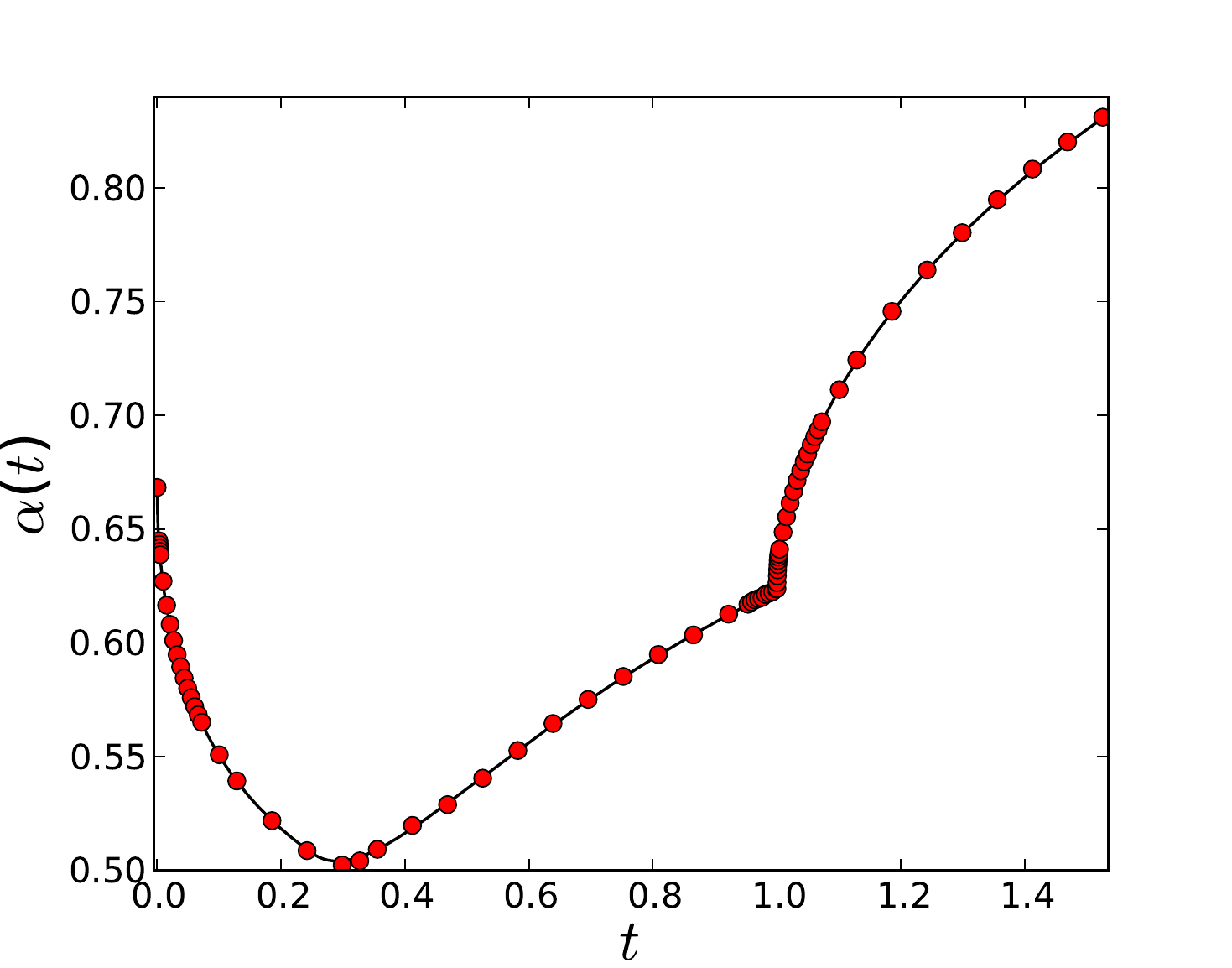}}
  \subfigure[]{\includegraphics[width=5cm, height=5cm]{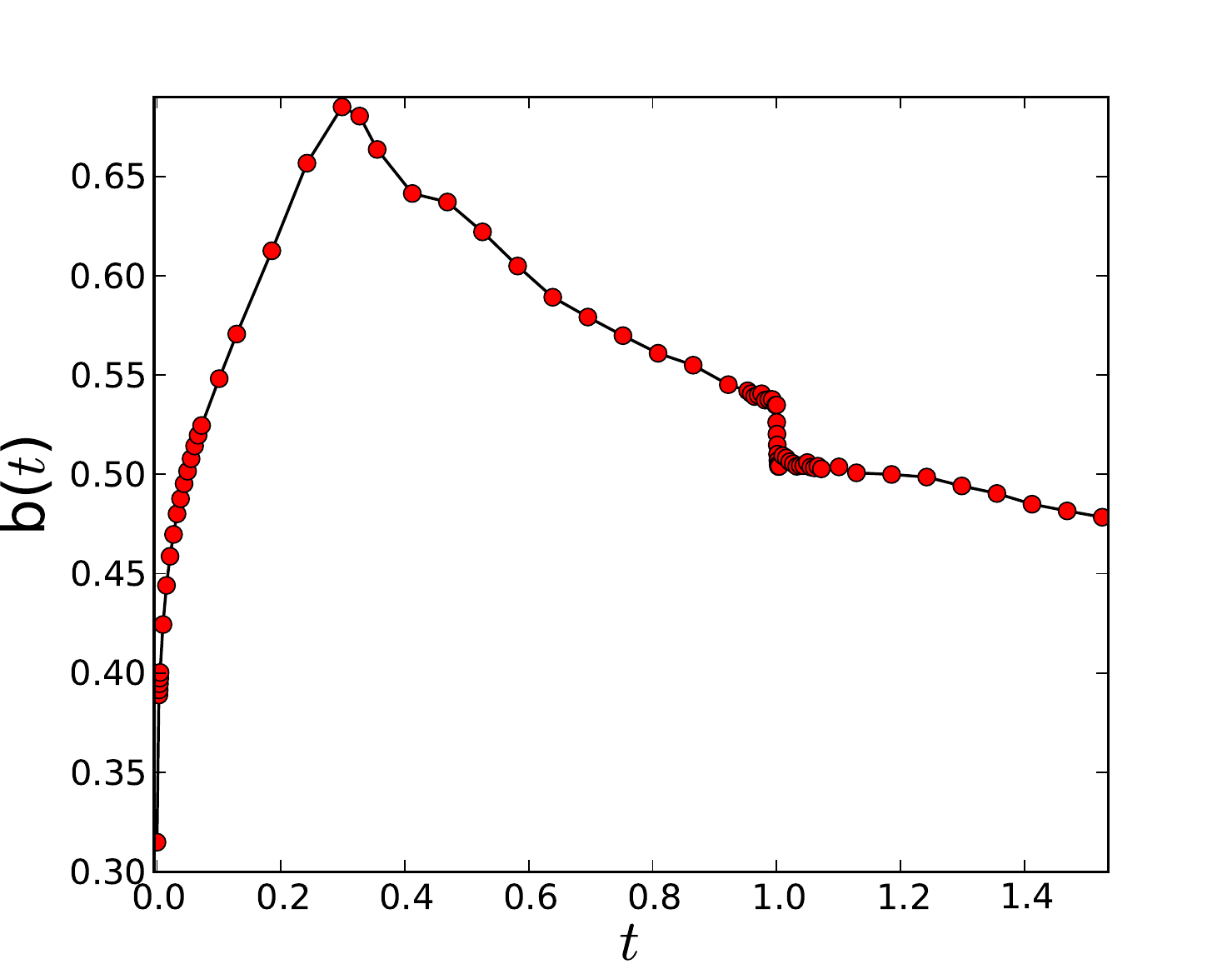}}
  \caption{(color online) (a) Numerically calculated entanglement as a function of 
the next-next nearest neighbor hopping $t$ for the dispersion in Eq. (\ref{eq:2d dispersion}) with $\mu=0$.
(b) The leading coefficient, $\alpha(t)$, as a function of $t$.  
(c) The sub-leading coefficient, $b(t)$, as a function of $t$. 
The solid line in (b) represents the analytical prediction from Eq.  (\ref{eq:alpha t 2d}).
}
\label{fig:2D}
\end{figure}

\begin{figure}[h]
  \includegraphics[width=0.92\columnwidth]{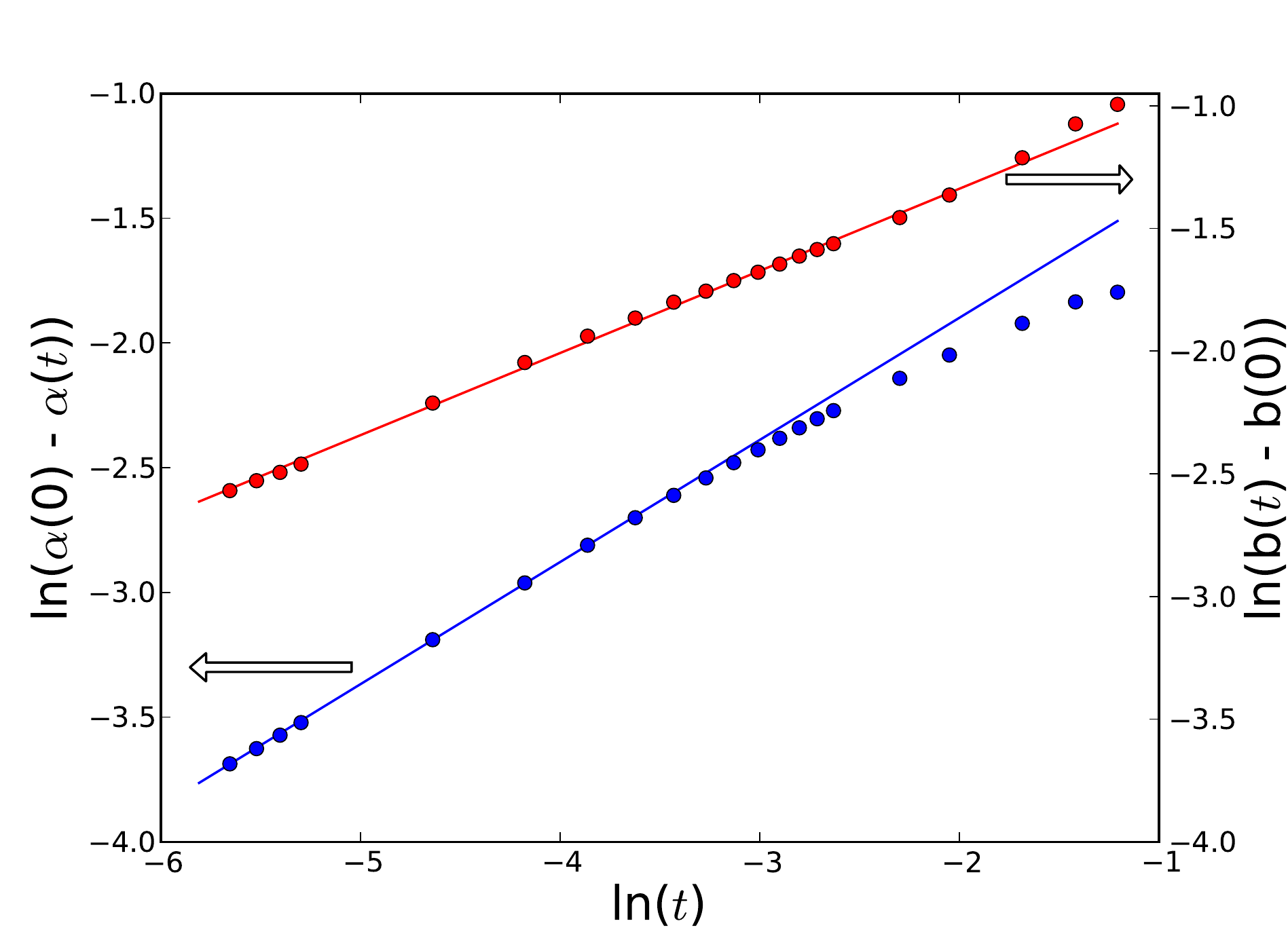}
  \caption{(color online) Log-log plot of $\alpha(0) - \alpha(t)$ (lower, blue line) and $b(t) - b(0)$ (higher, red line) against $t$. The parameter $t$ goes up to $\sim$ 0.3. The small $t$ behavior clearly shows a power law relationship for both as a function of $t$. The slope for the $\alpha(0) - \alpha(t)$ line is 0.490, whereas for the $b(t) - b(0)$ line it is 0.335. 
}
  \label{fig:2DfitDisp1}
\end{figure}

We now turn to the two-dimensional case.
The entanglement entropy is calculated for 
the tight binding model on the square lattice.
With the nearest-neighbor hopping whose magnitude is normalized to $1$
and the next-next-nearest neighbor hopping $t$, 
the dispersion is given by Eq. (\ref{eq:2d dispersion}).
The chemical potential is fixed to $\mu=0$, 
and $t$ is tuned to induce Lifshitz transitions.
Under the transformation $c_i \rightarrow c_i (-1)^{i_x + i_y}$,
the energy dispersion transforms as 
$\epsilon_t(k) \rightarrow - \epsilon_{-t}(k)$.
This implies that the entanglement entropy is an even function of $t$.
We therefore focus on the regime with $t \geq 0$.
Results are presented for subsystems of size $\ell$ = 5 up to $\ell$ = 80
with $0 < t < 1.5$. 
The entanglement entropy is expected to take the general form
\be
  \mathcal{S}(\ell) = \alpha(t) \ell \ln\ell + b(t) \ell + c(t) \ln \ell + d(t). 
\label{eq:form 2d}
\ee
In Fig.~\ref{fig:2D}(a), we show the entanglement entropy as a function of $t$ for $\ell=80$.
The peak at $t=0$ and the kink at $t=1$ become more pronounced as $\ell$ increases.
However, the overall shape of the curve is similar for different $\ell$.
In Figs.~\ref{fig:2D}(b) and \ref{fig:2D}(c), the first two leading coefficients are shown,
where the coefficients are obtained by fitting the entanglement entropy
as a function of $\ell$ in the form of Eq. (\ref{eq:form 2d}) upto $\ell = 80$.
The leading-order coefficient $\alpha(t)$ agree quite well with Eq.~\eqref{eq:alpha 2d} as predicted by Widom's conjecture.
In particular, $\alpha(t)-\alpha(0)$ is found to behave as $-s |t|^{r}$ near the critical point at $t=0$ with 
$s \approx 0.40 \pm 0.04$ and $r \approx 0.490 \pm 0.009$ as  shown in Fig. \ref{fig:2DfitDisp1}, fairly consistent with the analytic predictions $s = (4/3\pi) \approx 0.4244$ and $r=1/2$.
We can also extract the scaling behavior of the next leading-order coefficient.
It was found that $b(t)-b(0)$ also follows a power law for $t$ small as is shown in Fig. \ref{fig:2DfitDisp1}. 
In this case, we have $b(t)-b(0) \sim |t|^{r}$ with $r \approx 0.335 \pm 0.006$ near $t=0$.
Although the sub-leading term itself is non-universal,
the critical exponent that governs the scaling behavior near the Lifshitz transition 
may be universal.
It is of great interest to establish this rigorously in the future.
It is remarkable that the sub-leading coefficient vanishes with a smaller power
than the leading coefficient. 
The Lifshitz transition at $t=1$ shows similar behaviors except that the non-analyticity appears only for $t \geq 1$ as expected.

\begin{figure}[h]
  \centering
  \subfigure[]{\includegraphics[width=5cm, height=5cm]{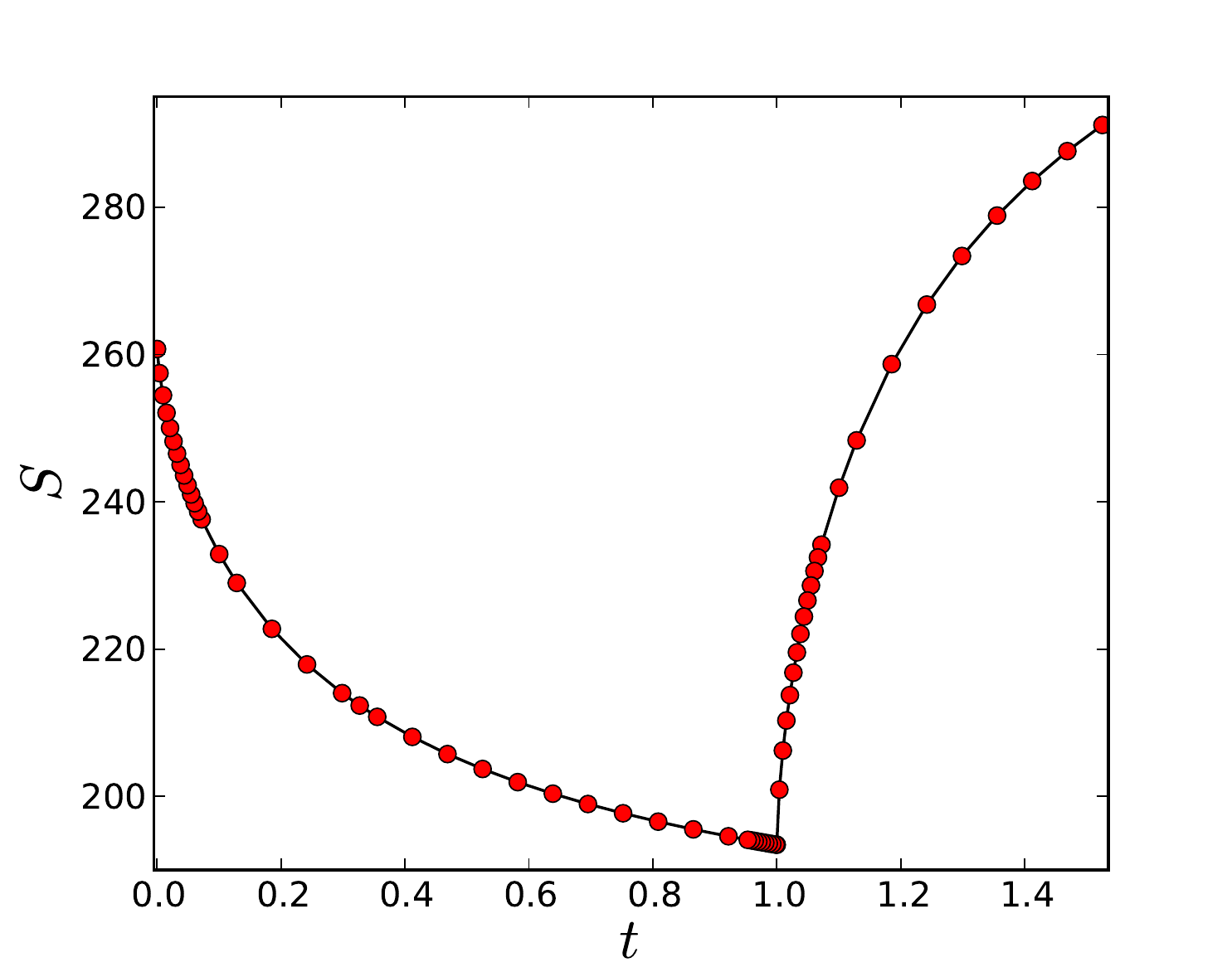}}
  \subfigure[]{\includegraphics[width=5cm, height=5cm]{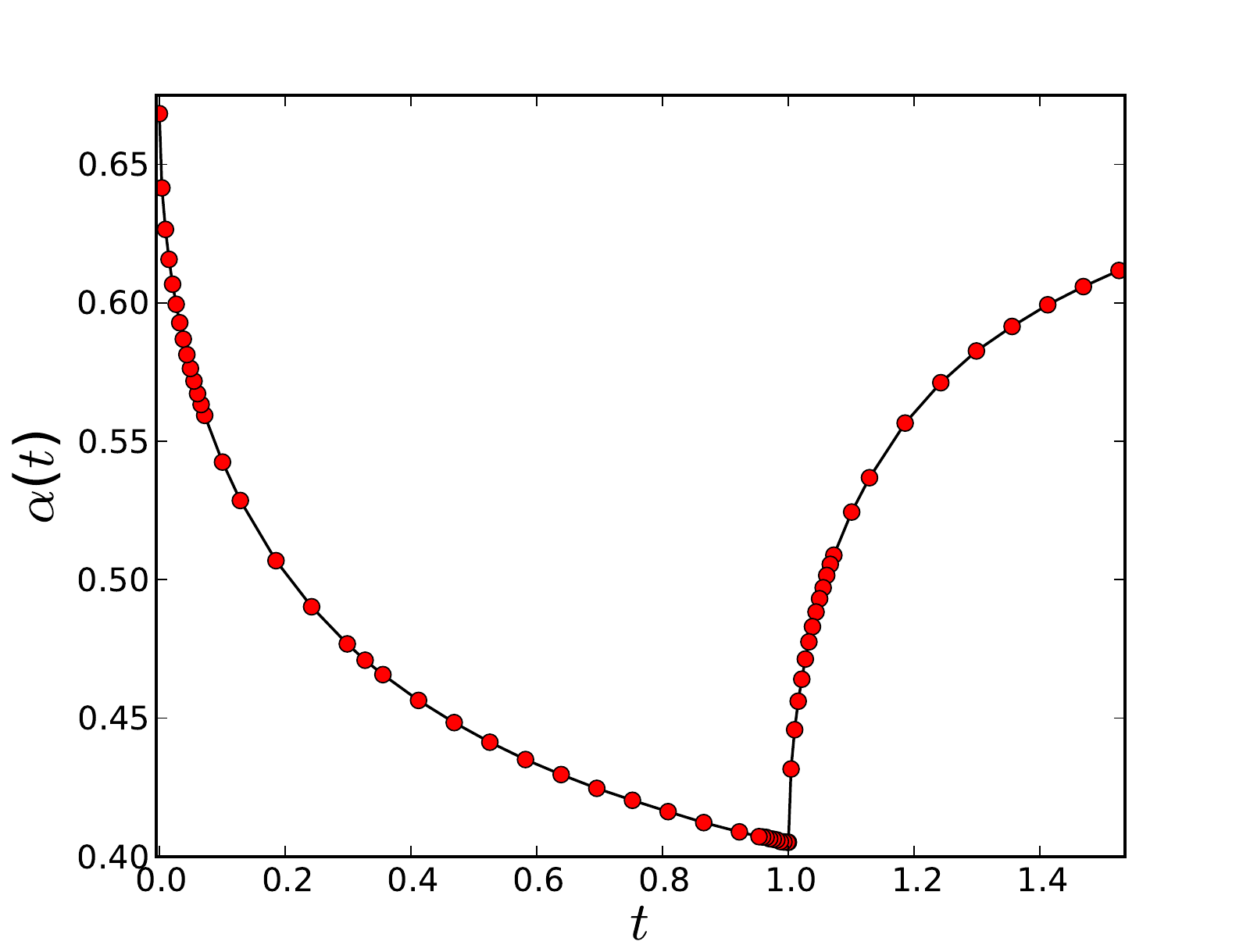}}
  \subfigure[]{\includegraphics[width=5cm, height=5cm]{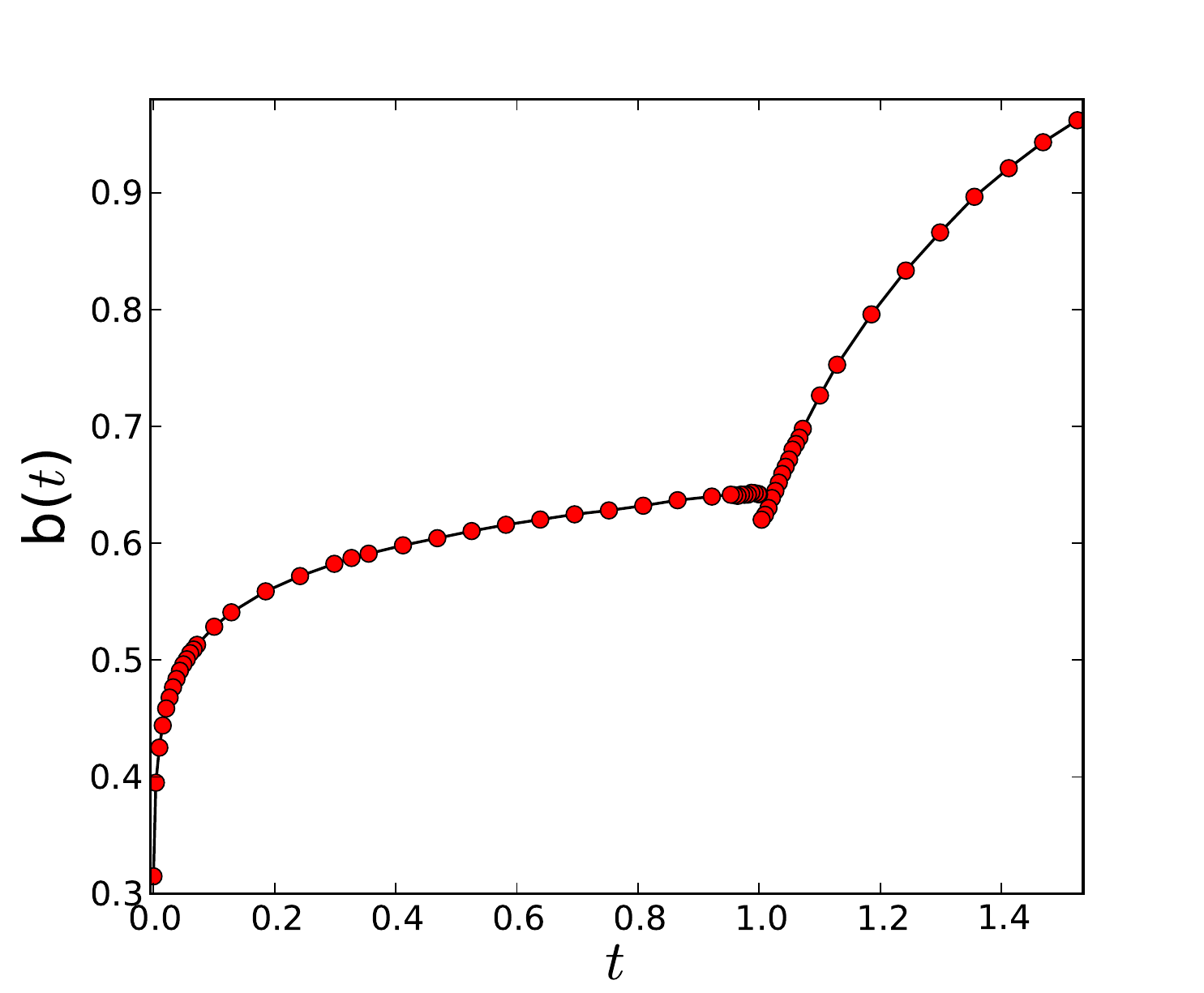}}
  \caption{(color online) (a) Numerically calculated entanglement as a function of 
the next nearest neighbor hopping $t$ for the dispersion in Eq. (\ref{eq:2d dispersion NN}) with $\mu=0$ and $\ell=80$.
(b) The leading coefficient, $\alpha(t)$, as a function of $t$.  
(c) The sub-leading coefficient, $b(t)$, as a function of $t$. 
}
\label{fig:2DNN}
\end{figure}

We also computed the entanglement entropy for the model with next nearest neighbor hopping 
with the dispersion
\be
 \varepsilon(\vk) = -2(\cos k_x + \cos k_y) - 2t [\cos (k_x + k_y) + \cos (k_x - k_y)].
\label{eq:2d dispersion NN}
\ee 
The entanglement entropy and the leading coefficients are displayed in Figs.~\ref{fig:2DNN}(a), \ref{fig:2DNN}(b), and \ref{fig:2DNN}(c). The behavior of $t$ near the critical values is similar to that which was shown for the previous dispersion.


\section{Effect of interactions}
How would the above conclusions for free fermions change in the presence of interactions? \emph{A priori}, interactions introduce new phases that are not amenable to the above analysis \cite{Meyer07,LeHur09, PhysRevB.79.205112}. In principle, the generic Lifshitz critical point may disappear due to a pre-emptive spontaneous symmetry breaking \cite{Kee03,Carr}.
In this case, the Lifshitz critical point becomes a multi-critical point that requires tuning more than one parameter.
At sufficiently low energies, however, the low energy modes are 
still described by a collection of one dimensional chiral fermions.
In light of the argument in Ref.~\cite{Swingle10}, 
the main features of the scaling of the entanglement entropy 
away from a Lifshitz transition in interacting systems with Fermi surfaces 
should be qualitatively similar for sufficiently large $\ell$. 
Indeed, the study of Lifshitz transitions in several mean-field approaches \cite{Yamaji06} suggests that the methods used in this work may be applicable to the quantitative analysis of such systems.
It would be of interest to examine the entanglement entropy
right at the critical point described by interacting 
quantum field theories with a general Lifshitz scaling, $z \neq 1$.

\begin{figure}[h!]
  \centering
  \subfigure[\ $V = 0.0$]{\includegraphics[width=0.49\textwidth]{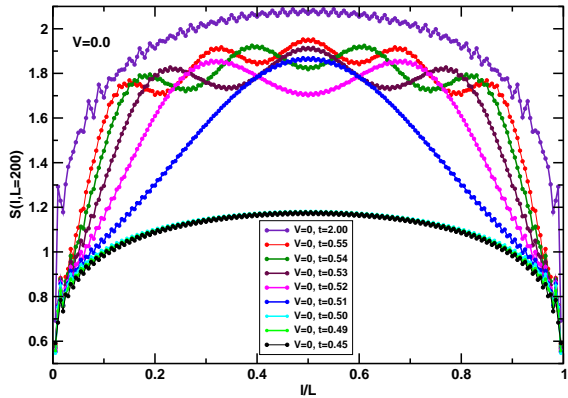}}
  \subfigure[\ $V = 0.2$]{\includegraphics[width=0.49\textwidth]{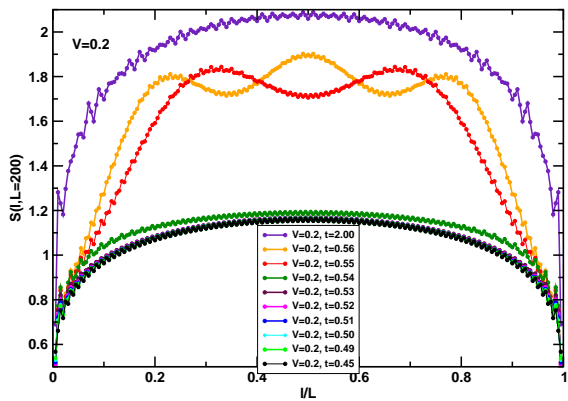}}
  \caption{(color online) DMRG results for Eq.~(\ref{eq:1dHV}) with $n=2$. 
    Results are shown for the bipartite entanglement entropy S(l,L) versus $l/L$ for systems with $L=200$ sites at half-filling with open boundary conditions.
    (a) Results for $V=0$, (b) results for $V=0.2$. }
  \label{fig:DMRG}
\end{figure}
In order to partly address some of these issues 
in one-dimensional case
we have performed DMRG calculations on the model
Eq.~(\ref{eq:1d Hamiltonian}) in the presence of an interaction term $V$:
\be
  \hat{H}_n = -\sum_i \left[
(\hat{c}^\dag_i\hat{c}^{}_{i+1} + t\hat{c}^\dag_i\hat{c}^{}_{i+n}) + \text{h.c.} 
+ V\hat{n}_i\hat{n}_{i+1} \right],
\label{eq:1dHV}
\ee
with $\hat{n}_i=\hat{c}^\dag_i\hat{c}^{}_{i}$.
A parallelized finite system DMRG method was used with the number of
states kept ranging from $m=256-512$ with open boundary conditions. By partitioning the system in left and right parts of linear extent $l$ and $L-l$, respectively, the  bipartite entanglement entropy $S(l,L)$ can then be
straight forwardly calculated. 
We first discuss our results for $V=0$, $n=2$ shown in
Fig.~\ref{fig:DMRG}(a). In contrast to results in the previous sections the DMRG calculations were performed at fixed particle number, at half-filling, as opposed to fixed chemical potential.
In this case the Lifshitz transition for $n=2$ occurs at $t=1/2$. We first focus on the results for $t=0.45, 0.49,0.50$ ($V=0$), which are nearly identical.
For gapless systems with open boundaries the uniform part of $S(l,L)$ is known from CFT~\cite{Calabrese04} to be
\be
S_u(l,L)=\frac{c}{6}\ln \left[\frac{2L}{\pi} \sin \left(\frac{\pi l}{L}\right)\right]+\mathrm{const},
\label{eq:cc}
\ee
with $c$ the central charge.
The uniform part of our results for $t=0.45, 0.49,0.50$ ($V=0$) follow this form very closely with a central charge of $c=1$. A known~\cite{Laflorencie06} alternating part of the entanglement entropy is also visible in the results
for $t=0.45, 0.49,0.50$ ($V=0$). When $t$ is increased beyond the Lifshitz transition at $t=1/2$ two effects are clearly visible: The entanglement entropy increases while pronounced oscillations develop. 
Such enhancement of entanglement entropy has been also observed in
other systems\cite{PhysRevB.79.205112,Essler}.
Since the system remains gapless
the uniform part of $S_u(l,L)$ should still be given by Eq.~(\ref{eq:cc}), but due to the pronounced oscillations in $S$ it is not possible to fit to this form until rather large values of $t$. 
At $t=2$ such a fit yields an
approximate central charge of $c\simeq 2$ corresponding to a doubling of the number of Fermi points. 
The wavelength of the oscillations correspond to the momentum associated with the small Fermi surface 
in Fig.~\ref{fig:fs1d-n2}(c). 
We expect similar oscillations to be observable in many other observables. 
Similar oscillations in the entanglement entropy have also been observed in spin chain systems.~\cite{White02,Andreas,Song12}.

Our results for $V=0.2$  $n=2$ are shown in Fig.~\ref{fig:DMRG}(b). As was the case in the absence of interactions the uniform part of the entanglement entropy is again described by Eq.~(\ref{eq:cc}). This continues to be the
case until $t\sim 0.54$, at which point the entanglement entropy again dramatically increases and oscillations appear. From these results for $S(l,L)$ no sign of another phase appearing is apparent and the only effect of the
interactions is to slightly shift the transition point away from $t=1/2$. 

We have also obtained preliminary results for $n=3$. In this case the expected tripling of the entanglement slows the DMRG calculations since the number of states that must be kept is significantly larger. 
However, our preliminary results indicate similar oscillations as for $n=2$ 
with the Lifshitz transition again preserved in the presence of weak interactions.
It would be very interesting to extend the DMRG results presented here in order to 
examine the possibility of a violation of logarithmic scaling 
at Lifshitz critical points with interactions present. 
We are currently investigating this issue.

\section{Conclusion}
We have theoretically investigated the scaling of entanglement entropy across Lifshitz transitions in gapless fermions 
where the transition is driven by a change in the topology of the Fermi surface. 
In one-dimension, the prefactor of the leading contribution to the entanglement entropy in the long distance limit, 
which is known to exhibit a logarithmic correction, 
jumps as the number of Fermi points increases across Lifshitz transitions.
The diverging length scale associated with the small Fermi pockets near the critical points
shows up as a crossover scale in the entanglement entropy.
In two-dimensions, it is shown that the leading and sub-leading coefficients of the entanglement entropy
exhibit scaling behaviors with distinct exponents. 
Preliminary DMRG results for the one dimensional case indicate that similar 
behavior is present in weakly interacting systems.

\section*{ACKNOWLEDGMENTS}
The work by HFS and KLH was supported by NSF Grant No. DMR-0803200, the Yale Center for Quantum Information Physics (DMR-0653377). 
The work by MR and SSL was supported in part by NSERC
and ERA from the Ontario Ministry of Research and Innovation.
The work by ESS was supported by NSERC.
Research at the Perimeter Institute is supported 
in part by the Government of Canada 
through Industry Canada, 
and by the Province of Ontario through the
Ministry of Research and Information. 
We also acknowledge discussions with Stephan Rachel, Nicolas Laflorencie, Israel Klich, Leon Balents, Matthew Fisher and Lesik Motrunich.
KHL and SSL thank KITP for the hospitality.


\end{document}